\documentclass[preprint,5p,times,twocolumn]{elsarticle}

\usepackage{amsmath}
\usepackage{mathtools,amssymb}
\usepackage{color}
\usepackage{xcolor}
\usepackage{graphicx}
\usepackage{caption}
\usepackage{subcaption}
\usepackage{caption}
\usepackage{enumitem, array}
\usepackage{bicaption}
\usepackage{dblfloatfix}
\usepackage{hyperref}
\usepackage{comment}
\usepackage[title]{appendix}
\usepackage[none]{hyphenat}
\usepackage{eurosym}
\usepackage{lipsum}

\hypersetup{
          colorlinks=true,          
          linkcolor=blue,           
          citecolor=blue,           
          filecolor=magenta,          
          urlcolor=blue,
          bookmarksdepth=4
      }

\usepackage{floatrow}
\usepackage{booktabs}                     
\usepackage{multicol}                     
\usepackage{multirow}                     
\usepackage{pdflscape}
\usepackage{siunitx}
\usepackage{makecell}
\usepackage{xltabular}
\usepackage{ltablex}
\usepackage{ltxtable} 

\usepackage{lipsum}
\usepackage{tabularx} 

\newcommand{\otoprule}{\midrule[\heavyrulewidth]} 

\newcolumntype{M}[1]{>{\centering\arraybackslash}m{#1}}

\usepackage{blindtext}

\journal{Materials Today}
\biboptions{sort&compress}

\begin{document}

\begin{frontmatter}

\title{3D printed architected lattice structures by material jetting}


\author[1]{Samantha Mora}
\author[1,2]{Nicola M.\,Pugno}
\author[1]{Diego Misseroni\corref{cor}}

\address[1]{Laboratory for Bioinspired, Bionic, Nano, Meta Materials and Mechanics, Department of Civil, Environmental and Mechanical Engineering, University of Trento, Italy}
\address[2]{School of Engineering and Materials Science, Queen Mary University of London, Mile End Road, London E1 4NS, UK}
\cortext[cor]{Corresponding Author (\it diego.misseroni@unitn.it) }


\begin{abstract}
High-precision 3D printing technology opens to almost endless opportunities to design complex shapes present in tailored architected materials. The scope of this work is to review the latest studies regarding 3D printed lattice structures that involve the use of photopolymers fabricated by \textit{Material Jetting} (MJ), with a focus on the widely used Polyjet and MultiJet techniques. The main aspects governing this printing process are introduced to determine their influence during the fabrication of 3D printed lattices.  Performed experimental studies, considered assumptions, and constitutive models for the respective numerical simulations are analyzed. Furthermore, an overview of the latest extensively studied 3D printed architected lattice materials is exposed by emphasizing their achieved mechanical performances through the use of Ashby plots. Then, we highlight the advantages, limitations, and challenges of the material jetting technology to manufacture tunable architected materials for innovative devices, oriented to several engineering applications. Finally, possible approaches for future works and gaps to be covered by further research are indicated, including cost and environmental-related issues.  
\end{abstract}


\end{frontmatter}


\tableofcontents
\section{Introduction}\label{intro}
\begin{figure*}[!h]
	\centering
	\includegraphics[width=1\textwidth]{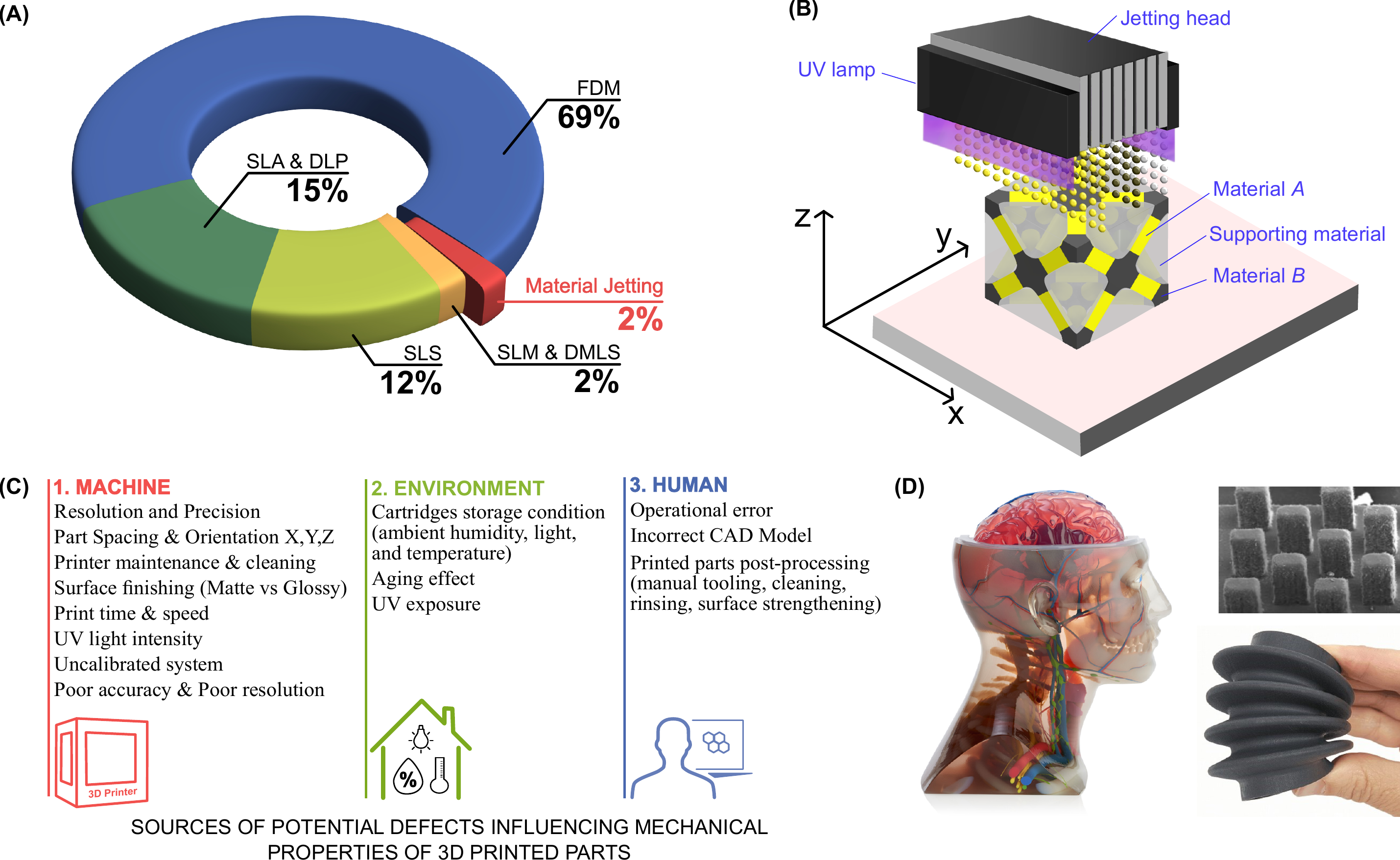}
	
	\caption{(A) Most used 3D printing techniques in the AM market for product and/or service applications~\citep{Statista}. (B) MJ general components and operating scheme. (C) Principal sources that affect the mechanical properties of final MJ printed parts. (D) Examples of 3D printed MJ parts with a variety of materials and colours (e.g. anatomical models. Reproduced with permission of~\citep{Stratasys2}. Copyright 2022 by Stratasys. All rights reserved. Microfluidic device~\citep{3DsystemsPrototyping}. Rubber-like spring. Reproduced with permission of ~\citep{Stratasys1}. Copyright 2022 by 2022 Stratasys. All rights reserved).}
	\label{fig:01}
\end{figure*}
Additive Manufacturing (AM) technologies produce three-dimensional free-form shapes and structures with complex geometries. The target design is sliced and then constructed layer by layer through 3D printers starting from a CAD file. Different materials and machines can be used depending on the selected technique existing in the current AM market. In the last years, the most demanded 3D printing technologies are Fused Deposition Modeling (FDM), Stereolithography (SLA), Selective Laser Sintering (SLS), Selective Laser Melting (SLM) and Material Jetting~\citep{Statista}, as it is shown in Fig.~\ref{fig:01}(A). The latter has been in a growing development since two decades ago,  because it allows the use photopolymers able to produce high quality and versatile components, with rigid, flexible, or temperature resistant characteristics~\citep{Stratasys2,3DsystemsPrototyping,Stratasys1}. In general, the most recent products fabricated by 3D printing are mainly oriented to the aerospace, automotive, biomedical, and robotics industries~\citep{daminabo2020fused}.
There are still sectors in the early stages, such as building construction, because fabrication at a large scale is still limited. Nevertheless, cutting-edge examples like 3D printed bridges~\citep{gardner2020testing} are leading to a growing expansion of AM. 
Materials development is one of the most remarkable applications potentiated by AM emerging technologies. Nature’s concepts related to placing energy and materials only where is necessary for specific structural functions (e.g. bones, teeth) have been a paradigm for the traditional rigid materials design. Consequently, breakthrough topologies, such as cellular and architected lattice materials, have been envisioned to prevent breakage but not losing strength and stiffness at the lowest weight possible~\citep{vogel1995better}.
\begin{table*}[h!]%
	\centering
	\caption{Overview of the main 3D printing processes, materials and resolution. Adapted from ~\citep{tofail2018additive}. The listed acronyms mean: SLA (Stereolithography), DLP (Direct Light Deposition), CLIP (Continuous Liquid Interphase Printing), SLS (Selective Laser Sintering), DMLS (Direct Metal Laser Sintering), SLM (Selective Laser Melting), EBM (Electron Beam Melting), FDM (Fused Deposition Modelling), LOM (Laminated Object Manufacturing), UAM (Ultrasonic Additive Manufacturing), SLCOM (Selective Lamination Object Manufacturing). Printing resolution refers to the expected layer thickness.}
	 {\renewcommand{\arraystretch}{1.3}
	\begin{tabular}{*{5}{>{\centering\arraybackslash}m{0.15\textwidth}}} 
		\otoprule
 \textbf{3D printing process} & \textbf{Main principle} & \textbf{Materials}& \textbf{Popular techniques} & \textbf{Printing resolution}\\
 \hline
Vat Polimerization & Polimerization & Rubber resin, Polymers & SLA, DLP, CLIP & 10 $\mu$m $^*$DLP:35-100\\
Binder jetting & Inkjet, Binder & Ceramics, Metals & Binder Jetting & 5–200 $\mu$m\\
\textbf{Material jetting} & \textbf{Inkjet, UV curing} & \textbf{Rubber resin, Polymers} & \textbf{Polyjet, Multijet} & \textbf{13–16 $\mu$m}\\
Powder bed fusion & Melting, Freezing powder & Metals, Ceramics, Polymers & SLS, DMLS, SLM, EBM & 80-250 $\mu$m\\
Direct energy deposition & Melting, Freezing & Metals, Ceramics &  Direct energy deposition & 250 $\mu$m\\
Material Extrusion & Melting, Freezing filaments &  Polymer, Ceramics & FDM & 50-200 $\mu$m\\
Sheet Lamination & Joining &  Metals, Polymer, Ceramics, Paper & LOM, UAM, SLCOM & variable \\
 \hline
	\end{tabular}}
	\label{Table:1}
\end{table*}
The latter also represents significant economic, energy savings for industrial purposes and  thus drastically reducing the environmental impact.  Nowadays, challenging lattice structures, meta and multi-materials with new capabilities have been easily fabricated by 3D printing. Thus, the limitations of common manufacturing technologies (e.g. cast molding, CNC) in time, materials selection, and geometrical complexity can be overcome. Moreover, stimuli-responsive and smart materials have started to be also conceived to add the transformation over time, the 4$^{th}$ dimension, to the 3D printing technique~\citep{rafiee2020multi}. Then, AM also opens a promising and broad research objective towards the fabrication of innovative and active parts.\\
The scope of the present review is to discuss the main outcomes and challenges of the latest research works related to architected lattice materials fabricated via \textit{Material Jetting} (MJ), with the main focus on commercial techniques that work with photopolymers, such as Polyjet$^\text{\textregistered}$~\citep{Stratasys1} and MultiJet$^\text{\textregistered}$~\citep{3DSystemsintro} (Fig.~\ref{fig:01}). High resolution, the use of rigid and soft curable resins in a single print, inherent morphing capabilities, and simplified post-processing activities have motivated the use of MJ for manufacturing novel lattice structures during the last years. Various examples have also been extended towards practical applications (e.g. energy absorption and impact devices, scaffolds, soft robotics). Therefore, it is important to analyze what are their achieved mechanical performances and to explore fabrication aspects that can limit the design targets. 
This review is organized in different chapters as follows. In Section \ref{matjetting}, the main aspects that govern the MJ process are briefly introduced, as well as the advantages and existing drawbacks with their potential solutions (e.g. curing rate, sustainability). Section~\ref{3DParchlat} starts with a general overview of architected materials. Then, the use of MJ in the fabrication of lattice structures, applied constitutive models, and the printing process effects on their resulting mechanical performances are discussed. Next, the following subsections present examples of 3D printed structures divided into two main categories, based on the constituent materials with their respective characteristics and functions, namely Single-material printed lattices (Rigid based, Soft and instability control, Auxetic and shape programmable, Wave propagation control) and Multi-materials printed lattices (Reinforced composites with rigid lattices, Different stiffness components, Functionally graded and active). 
In Section~\ref{chall}, we analyze the challenges and limitations concerning the MJ printing process under the lattice manufacturing perspective and further experimental work required (e.g. anisotropy assessment, dynamic and fatigue tests). In addition, geometrical limitations in regard to MJ manufacturing and a cost estimation related to actual prices are presented. Furthermore, we quantitatively evaluate  the actual outcomes of the exposed MJ architected lattices by comparing their achieved mechanical performances with respect to already existing materials in the Ashby plots (e.g. Young's modulus vs. density). Finally, in Section~\ref{conclusions}, we present concluding remarks  by pointing out research gaps and future directions that lead to new architected materials with functional, high-performance, and innovative applications by exploiting MJ capabilities.

\section{Material Jetting (MJ) technology overview}\label{matjetting}
The existing AM techniques, classified according to ASTM-ISO standards, work with different processes, types of raw materials, and printing resolution~\citep{tofail2018additive}, as it is summarized in Table~\ref{Table:1}. 
Material extrusion techniques, such as Fused deposition modeling (FDM), are more widely used in the AM market than Vat photopolymerization or Direct Deposition due to the cost convenience, scalability, variety of materials ranging from polymers and gels to bio-based~\citep{daminabo2020fused,Statista,Sculpteo}, as it is shown in Fig.~\ref{fig:01}(A). When higher quality is required, MJ printing allows obtaining polymeric objects owning high dimensional accuracy (layers height 13-16 microns, accuracy of $\pm$0.06-0.1\% for part length under $\approx$100~mm) with less surface roughness, reduced staircase defects, and less waste with respect to material extrusion process~\citep{Stratasys1,gulcan2021state,3DsystemsBrochure}. Thus, MJ is a promising technique for the AM field. However, the maximum size of the printed object is limited by the actual tray size (to date $<$ 1 m), and the minimum wall thickness should be $>=$ 0.6 mm since tiny parts tend to be fragile and warped, especially during post-processing operations (see details in Section~\ref{man:limits}).

MJ printers use multiple nozzles that deposit base and support material droplets concurrently. A roller and a wiper unit level the printed layers to avoid excess material, as it is schematized in Fig.~\ref{fig:01}(B). Further details on the MJ technique are referred in Appendix~\ref{app_MJprocess}. Simultaneously, a curing process with Ultraviolet (UV) light is carried on for hardening the deposed layers, and thereby, the final part is formed. Radical, cationic, or hybrid photocuring mechanisms can be involved~\citep{quan2020photo}. Higher levels of UV energy influence the glass transition temperature ($T_{g}$) of the resin, due to an increasing degree of polymerization of the material and its cross-linking density~\citep{zabti2012effects}. Therefore, photopolymers can reach a viscoelastic or rubbery behavior at low temperatures. Typical MJ resins exhibit a $T_{g}$ of about 50$^{\circ}C$. Thus, viscoelastic properties might be observed on printed parts at ambient temperature~\citep{blanco2014nonisotropic}. The orientation of the part on the build-tray affects the relaxation modulus E(t), which also varies in time ~\citep{gay2015analysis}.
 The photopolymerization of the resins requires to be done at a fast rate to guarantee its optimal solidification. Photopolymers with low viscosity (0.1-10 Pa.s) are ideal because they facilitate their flow through the heated nozzles and their spreading onto the build tray. However, the balanced weight of the constituent oligomers and monomers influences the degree of crosslinking and viscosity of the resin~\citep{tan2020recent}. High-performance polyamide resins are an attractive candidate for fast curing rates at low viscosity, and at the same time offer high thermal stability, mechanical properties, and adhesion strength between layers~\citep{wagner2019cure}. See further details in Appendix~\ref{app_MJprocess}.

Although the most used MJ techniques (Polyjet$^\text{\textregistered}$~\citep{Stratasys1} and MultiJet$^\text{\textregistered}$~\citep{3DSystemsintro}) work similarly, the main difference relies on the type of support material and post-processing activities. In the case of Polyjet, soluble gel materials are utilized and they can be removed mechanically (by water pressure) or chemically (by a 2\% sodium hydroxide ($Na (OH)$) - based solutions enriched with the 1\% of sodium meta-silicate ($Na_{2}SiO_{3}$)). The latter method is preferable in the case of models with complicated geometries that are difficult to be cleaned and/or made up of very thin and fragile elements (e.g. lattice structures). During the removal operation, it is suggested a regular check of the printed parts. The solution temperature should not exceed 30°C, and the printer guidelines suggest that soaking time should not last longer than one hour to avoid unwanted deformation, especially for models with thin walls (less than 1~mm). The complex geometry typical of 3D printed architected lattices might require a longer soaking time for complete support removal. Nevertheless, we have observed that the support material is melted almost completely in less than 4 hours, even for very detailed architectures with hard-to-reach spots (in the case of lite support grids and using a clean station). See also Appendix~\ref{app_MJprocess}. In this case, a constant visual inspection of the parts is necessary to control the integrity of the thin elements. Once the cleaning process is finished, the parts should be rinsed in water to remove any residual supporting material. Finally, it is suggested dipping the cleaned parts into a glycerol-based solution for 30-60 seconds to strengthen the surfaces and limit aging effects~\citep{StratasysSUP,Nahumsupport}. MultiJet employs bulk wax-based supports. They can be removed first by heat sources above 65°C for 30~minutes and longer. The removal of fine remainders on the 3D-printed parts requires solvent baths (EZ Rinse-C or light mineral oil) in shorter periods (from 2 to 5 minutes). Water at a temperature between 30-35°C with soap is recommended to rinse the parts~\citep{3Dsystemsclean}.
\begin{figure*}[!h]
	\centering
	\includegraphics[width=1\textwidth]{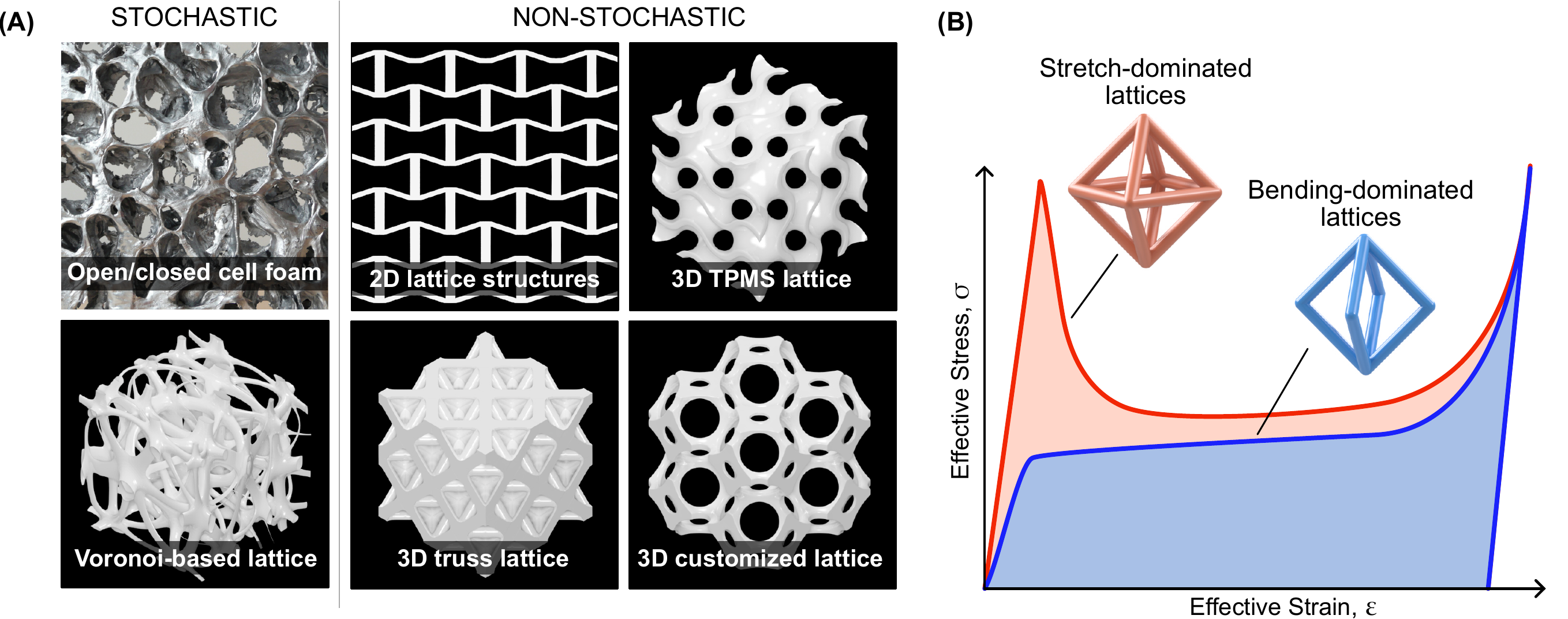}
	\caption{(A) Cellular and lattice structures general classification. (B) Example of characteristic effective stress-strain curves $(\sigma-\epsilon)$ of stretch and bending dominated lattices, inspired by~\citep{ashby2006properties}.}
	\label{fig:02}
\end{figure*}
MJ uses simultaneously various polymers with different stiffness, physical properties, or color addition in a single print, as it is illustrated in Fig.~\ref{fig:01}(D).
The most commercialized photocurable resins, summarized in Table~\ref{Table:3}, reach several values of elongation at break, close to 200\% for rubber-like and 10\% for plastics resins~\citep{Stratasys3,3DSystems}. See also Appendix~\ref{app_MJprocess:mech}. Thus, they can be applied in a variety of functional applications. Specific combinations among polymers and flexible rubbers can form a strengthened material within a wide shore hardness range (A30-A95), even though some resins may exhibit poor mechanical performances and degradation over time~\citep{Varotsis}.
Despite a still limited number of available material jetting resins, there are upcoming studies focused on creating new printing materials from non-photopolymers and low melting point metals, like aluminium~\citep{edgar2015additive}. The drawback of using current photocurable materials is that they are not \textit{eco-friendly} and not recyclable in general~\citep{shapira2017next}. For this reason, there is a growing interest to find less toxic and sustainable alternatives such as bio-based monomers~\citep{PIERAU2022101517}, bio-resins made of photocurable oligomers~\citep{voet2021sustainable}, and cell-laden cross-linkable photopolymers like hydrogels~\citep{ChiulanPhotopolymerization}. 

\section{3D printed architected lattice materials}\label{3DParchlat}
\subsection{Architected lattice materials overview}

Monolithic and traditional materials have been used for specific requirements of strength, stiffness, and toughness. The achievement of improved mechanical properties by saving weight and costs without compromising the structural integrity has motivated the continuous development of architected lattice materials for several engineering fields. According to the classification of the cellular structures~\citep{ashby1997cellular,tao2016design,tamburrino2018design}, shown in Fig.~\ref{fig:02}(A), lattice structures  can be either stochastic or non-stochastic. The former consist of random material distribution nowadays designed by a generative algorithm, while the latter are generated via tessellation of a unit cell formed by interconnected faces, edges, or struts/ties as trusses, usually linked by pinned or fixed nodes~\citep{ashby1997cellular}. These complex geometries with optimized topologies are able to withstand high loading values and large deformations at low density values. As metamaterials, architected lattice structures can exhibit negative Poisson’s ratio \textit{(NPR)}~\citep{saxena2016three,lakes2017negative,kolken2017auxetic}, negative coefficient of thermal expansion \textit{(NCTE)}~\citep{wang2016lightweight,wu2019mechanical,lim20192d}, or negative refraction index~\citep{smith2004metamaterials,PADILLA200628,Shalaev2007,PhysRevLett.102.023901}. By modifying the material configuration and microarchitecture other novel properties can be obtained such as waves attenuation, acoustic insulation, and energy harvesting~\citep{valdevit2016fabrication,jia2020engineering,wu2019mechanical}. New targets can also be envisioned, for example, simultaneous load-bearing and thermal capabilities, stiffness and impact resistance~\citep{valdevit2018architected,woesz2004cellular,fleck2010micro}, or \textit{NPR} and \textit{NCTE} ~\citep{ai2017metamaterials,ai2018three}.
Lattice material properties are controlled by relative density, nodal connectivity, effective Young’s Modulus (EEM), yield, and fracture strength~\citep{tao2016design}. Relative density is the ratio between the densities of the lattice material and its solid counterpart~\citep{fleck2010micro}. 
For 2D lattices with $t/l<<$ the relative density can be calculated by the formula $\rho_{rel}=A t/l$, where \textit{t} is the ligament thickness, \textit{l} is the ligament length, and \textit{A} is a coefficient of proportionality depending on the topology of the considered lattice~\citep{ashby1997cellular,fleck2007damage}\footnote{For 2D hexagonal and triangular lattices $A=2\sqrt3$, while for Kagome lattice $A=\sqrt3$.}. For 3D open-cell foams the relative density $\rho_{rel}$ scales with  $(t/l)^2$, while for 3D closed-cell foams it scales with $t/l$. For lattice truss materials with a radius size r, the relative density is defined as $ \rho_{rel}=(6\sqrt{2}+2)\pi(r/l)^2$~\citep{fan2008yield}.
The level of nodal connectivity determines the predominant behavior of a lattice structure and the collapse mechanism~\citep{deshpande2001foam,kudo2019compressive}.  \textit{Stretch-dominated} lattices present high connectivity level and present high initial strength and stiffness~\citep{deshpande2001foam}. Their collapse mechanism can be plastic stretch-dominated, buckling-dominated, stretch–fracture-dominated~\citep{ashby2006properties}.
On the other hand, the \textit{Bending-dominated} lattices are characterized by low nodal connectivity. The fact of presenting less initial strength but attained high strain values, makes them suitable for energy absorption applications. Moreover, they exhibit a regular and extended stress plateau after yielding, with large deformations at low stress values. Their failure mechanisms can be bending-dominated, buckling-dominated, and fracture-dominated~\citep{ashby2006properties}. Both lattice behaviors, stretch and bending dominated, can be distinguished in the different stress versus strain curves from Fig.~\ref{fig:02}(B).

\subsection{Material jetting process effects on mechanical properties of 3D printed lattices}

The latest developments in 3D printing techniques have allowed the possibility to produce more complex and enhanced architected materials, inspired by mathematical (e.g. triply periodic minimal surfaces, fractals) and biological mimicry models (e.g. honeycombs, glass-sponge inspired) with outstanding mechanical performances~\citep{ambekar2021topologically}. MJ appears as a promising technique to fabricate lattices  from mm-scale to cm-scale due to the high resolution it offers, the use of single and/or multi-materials, and less manual post-processing. 

In every 3D printing process, there is a permanent target to assess the influence of manufacturing parameters on the geometry accuracy and performance of the final printed parts. In the case of MJ, the build part orientation, UV curing, supports removal operations besides other additional factors, summarized in Fig.~\ref{fig:01}(C), have effects on the tensile strength and modulus of 3D printed specimens \citep{barclift2012examining, dizon2018mechanical} (see further details in Appendixes~\ref{app_MJprocess:mech} and~\ref{app_MJ_printing:effects}).

\begin{figure*}[!h]
	\centering
	\includegraphics[width=1\textwidth]{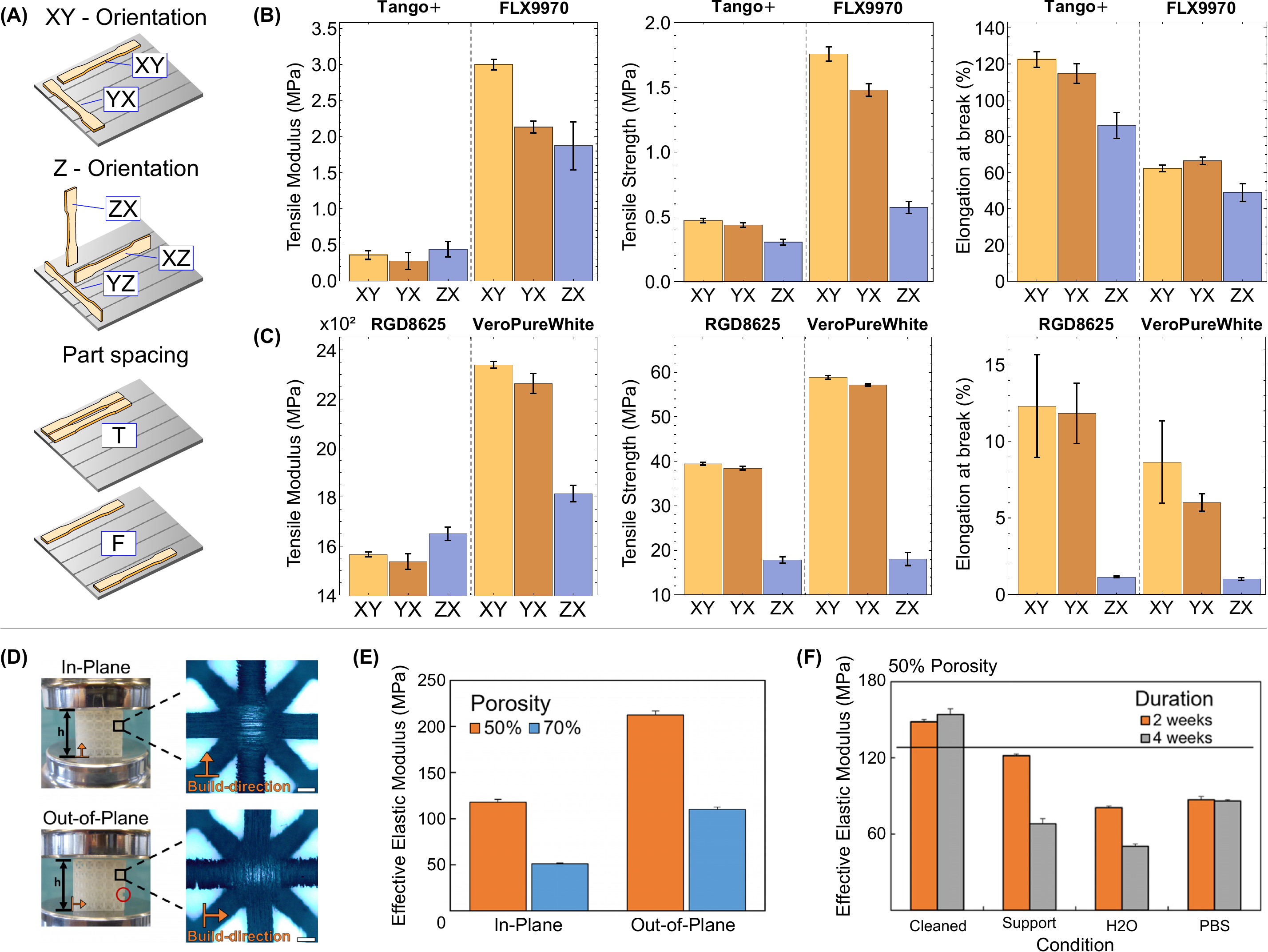}
	\caption{(A) Reference sample orientation and spacing (Tight or Far) on the build tray. Tensile modulus, tensile strength and elongation at break exhibited by specimens with different print orientation, made of (B) Flexible materials (Tango+ and FLX9970), and (C) Rigid materials (RGD8625 and VeroPureWhite). Adapted from~\citep{konigshofer2021mechanical}. (D) Lattice samples tested “In-plane” and “out-of-plane” orientations with respect to the compressive load application. Variation of Effective elastic modulus of lattices under compression by considering: (E) Diverse porosity and build orientation and (F) Environmental storage conditions. Reproduced with permission of \citep{egan2019mechanics}. Copyright 2019 ASME.}
	\label{fig:03}
\end{figure*}

A common source of anisotropy relies on the layers' print direction and adhesion, or surface roughness. MJ fabricates parts with a level of anisotropy of approximately 2\%, which is lower than other usual commercial AM techniques that reach even 50\% (e.g. FDM)~\citep{kazmer2017three}. (Details in Appendix~\ref{app_MJ_printing:effects} and~\ref{app_MJ_printing:effects}). During the photopolymers deposition, the previous layers are compacted by the upper resin layer. Moreover, they are subjected to subsequent rounds of UV curing at gradually less energy, since the UV light has to penetrate through an increasing number of stacked layers. This evolutionary process leads to obtaining overall cured and well-bonded layers. Therefore, the resulting objects present very homogeneous mechanical and thermal properties.   However, it must be acknowledged that parts built parallel to the print tray ($XY$ direction), as shown in Fig.~\ref{fig:03} (A), presented the highest tensile strength, tensile modulus, and elongation at break values.  While those in perpendicular orientation had the lowest mechanical resistance~\citep{barclift2012examining,maroti2018printing,bochnia2016anisotrophy,ulu2015enhancing,konigshofer2021mechanical,das2018effect}. This fact is illustrated in Fig.~\ref{fig:03} (B) and (C) from tensile tests carried on specimens printed either with flexible or rigid materials~\citep{konigshofer2021mechanical}.  An  exception is observed on samples printed with flexible Tango+ and RGD8625 (combination of rigid VeroClear and soft Tango+). In these cases, the highest values of tensile modulus are obtained on parts built perpendicular to the print tray ($ZX$ direction), as shown in Fig.~\ref{fig:03} (B) and (C). Moreover, less spacing among the printed parts on the build platform (see Fig.~\ref{fig:03} (A)) leads to statistically significant improvements on the mechanical properties because of a longer UV light curing exposure~\citep{barclift2012examining}. Although the UV curing time may be longer in Z-oriented specimens, they contain a higher number of printed layers with a minimum level of cohesion between the previous and the upcoming ones. Then, a consequent weakening effect is expected~\citep{mueller2015tensile,cazon2014polyjet}. 
This fact can also be seen in lattices, whose Young's modulus becomes variable and tends to decrease if the printed layers were oriented "perpendicularly" to the compression load (in-plane case)~\citep{egan2019mechanics,mueller2018buckling,dalaq2016mechanical}, as displayed in Fig.~\ref{fig:03} (D) and (E).
Furthermore, the rigid photopolymers used in MJ present levels of fracture strength that significantly depend on the print orientation~\citep{vu2018characterizing}. 
The numerical modeling of Polyjet and Multijet architected lattices requires different constitutive models depending on the printing photopolymer. Material characterization studies of their constituent photopolymers provide some indications on the best appropriate constitutive model to be considered, as summarized in Appendix~\ref{summary:CM:AL}, Table~\ref{tab:LatCModels}. Traditionally, MJ printed materials have been represented by bilinear~\citep{salcedo2018simulation} or viscoelastic orthotropic~\citep{blanco2014nonisotropic,vu2018characterizing} constitutive models. 
Thus, their time-dependent behavior and high nonlinear stress-strain relationships can be described. In the case of flexible materials, hyperelastic properties have been evidenced. Several strain energy density models have been developed for accurately fitting experimental data. For example, Ogden 4th-order for rubber-like~\citep{berselli2011hyperelastic}, whereas Arruda-Boyce ~\citep{tee2020polyjet}, Yeoh~\citep{slesarenko2018towards} and Mooney-Rivlin~\citep{salcedo2018simulation}, for soft digital materials. Moreover, a transversely isotropic hyperelastic-viscoplastic model for glassy photopolymeric lattices enables the integration of the anisotropy caused by the printing direction effects, with the expected inelastic deformation and strain rate. An additional modified Tsai-Wu failure criterion, that considers strain-softening for anisotropic materials, also complemented this approach~\citep{zhang2016transversely}.  In addition to the effects attributed to the printing process, there are post-processing stages that tend to affect the expected performance of 3D printed lattices fabricated via MJ (e.g. support removal, thermal treatments, storage conditions). For example, alternatives to simplify support removal operations have started to be investigated. In particular, a snap-fit fabrication method can reduce the amount of required support material and the printing time of body-centered cubic (BCC), face-centered cubic (FCC), and octet lattices. Groups of struts were printed separately and later connected by the nodes with a polymer adhesive. Mechanical strength and energy absorption values also presented a significant increment in the order of 100\% with respect to integrated lattices printed in one round~\citep{liu2020maximizing}. As a complement of a geometrical optimization approach, a thermal treatment after the printing stage produced improvements in the mechanical performances of lattices. Then, higher energy dissipation values, up to 60\%, under large compressive strains were obtained~\citep{chen20183d}. Furthermore, the external exposure conditions over time also influence the effective elastic modulus, as seen in Fig.~\ref{fig:03}(F). Cleaned lattice samples stored at room temperature  reached higher values with respect to samples placed in water, support material, and saline solution~\citep{egan2019mechanics}.

\begin{figure*}[!b]
	\centering
	\includegraphics[width=0.95\textwidth]{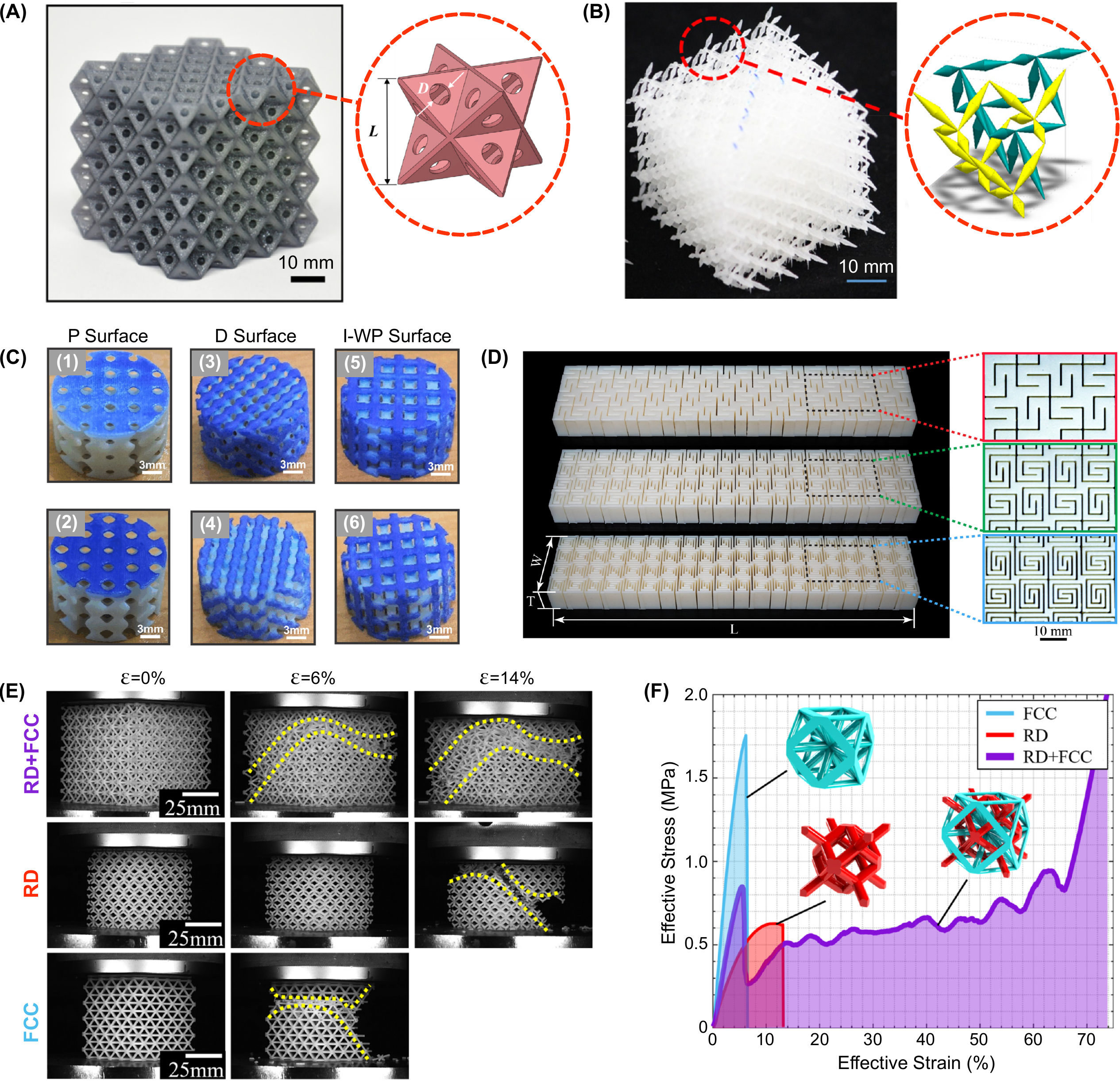}
	\caption{(A) Face-centered cubic semi-plate lattices. Reproduced with permission of~\citep{li2021additive}. Copyright 2021 Elsevier. (B) Hybrid pentamode. Reproduced with permission of~\citep{mohammadi2020hybrid}. Copyright 2020 AIP Publishing. (C) TPMS scaffolds with radially graded porosity with types P, D, I-WP. Reproduced with permission of~\citep{afshar2018compressive}. Copyright 2018 Elsevier. (D) Lattice metamaterials with one, two, and three orders of fractal cuts. Reproduced with permission of~\citep{zhang2021harnessing}, Copyright 2021 Elsevier. (E) Failure, shear bands, and strain development during compression test of single RD, FCC, and IPL lattices and (F) their respective stress-strain curves comparison. Reproduced with permission of~\citep{white2021interpenetrating} under Creative Commons Attribution 4.0.}
	\label{fig:04}
\end{figure*}



\subsection{Single material 3D printed lattices}

MJ has allowed the manufacturing of a variety of 3D printed architected lattice materials, from simple to challenging geometries, and achieving tailored mechanical properties (e.g. strength, toughness, auxeticity, instability, and wave control). The following subsections present notable examples of lattices fabricated from a single material, either rigid, flexible or \textit{hybrid}. Hybrid materials are composites obtained by combining fractions of rigid and soft MJ photopolymers, to obtain materials with different shore levels or functional-graded materials. Such hybrid materials adopt different commercial names, \textit{Digital material}, or \textit{Multi-material composites} in the case of Polyjet and MultiJet techniques, respectively (see also Appendix~\ref{app_MJprocess:mech}).

\subsubsection{Rigid-based material architected lattices }
Glassy photopolymers have been mostly considered and characterized to manufacture lattices due to their affordability with respect to other MJ photo-curable resins. 
The design of 3D printed lattices fabricated with a single, stiff material has usually included quasi-static experimental tests besides numerical simulations. Their principal target was to find improved mechanical performances by tuning geometrical features from diverse architected topologies. 
For example, tailored 3D printed honeycombs presented a variable cell thickness gradient in hexagonal, re-entrant, and chiral configurations. At lower gradation parameters, energy absorption efficiency grasped peak values around 90\% (approximately 65\% in experiments) versus 50\% observed in traditional materials~\citep{kumar2019tunable}.  High compressive strength was obtained by spherical lattices with optimized mass ratio~\citep{lancea2020compressive} and triply periodic minimal surfaces (TPMS based on Schoen's FRD and  OCTO lattices)~\citep{sathishkumar2020mechanical}. Alternatives like semi-plate lattices, as presented in Fig.~\ref{fig:04}(A), presented enhanced load-bearing capabilities compared to truss-based lattice materials. Likewise, they obtained similar fracture toughness and higher energy absorption values than metal foams, but at lower relative densities. The inclusion of holes on the plates’ surfaces allowed the control of crack propagation, and uniform stress distribution, besides an easier support material removal from inside~\citep{li2021additive}.
Moreover, pentamodes lattices illustrated in Fig.~\ref{fig:04}(B) were fabricated to assess the effect of varying their midpoint lattice connection. As it approximates the centroid of the respective unit cells, the elastic and shear modulus decreases, whereas the Poisson's ratio remains almost unchanged~\citep{mohammadi2020hybrid}. 
Furthermore, MJ has started to be widely used to fabricate prototypes for biomedical devices with sufficient strength and porosity, especially for bone or tissue growth. Different scaffold topologies  have been designed and tested, such as rectangular beam-based~\citep{egan2019mechanics}, triply periodic minimal surfaces (TPMS of the type P, D, I-WP, with uniform and graded porosity, as shown in Fig.~\ref{fig:04}(C))~\citep{afshar2018compressive, kadkhodapour2014investigating}, vintile~\citep{abate2020design}, cubic~\citep{alaboodi2018experimental}. The lattice architecture affects the deformation behavior, where the ligaments with heterogeneous mass distributions lead to the increment of stress concentrations. Scaffolds with higher porosity and lower ligament sections, exhibited decrements in elasticity and yielding strength For example, when shifting from relative density values of 60\% to 30\%  in TPMS (types P and D), Young's modulus tended to decrease approximately in a 30\% and yield stress was reduced in between 10-20\%~\citep{kadkhodapour2014investigating}. Moreover, translucent scaffolds obtained improved stiffness at higher porosity values than their colored counterparts. A more ductile failure was also observed due to the presence of opened-up crazes on the surfaces, instead of brittle surfaces with voids and craters that induce a faster crack propagation ~\citep{alaboodi2018experimental}.
\begin{figure*}[h!]
	\centering
	\includegraphics[width=0.95\textwidth]{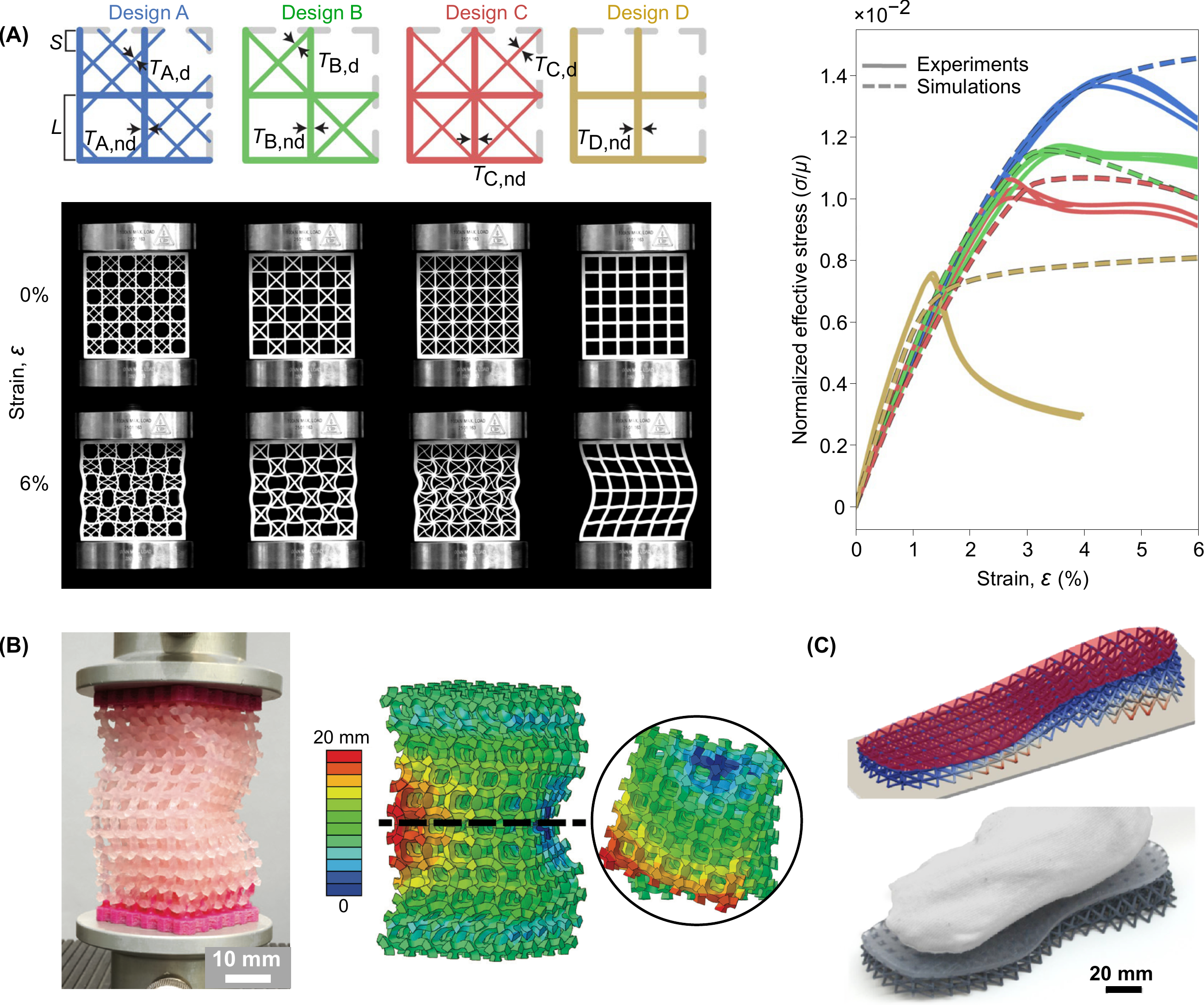}
	\caption{(A) Mechanical deformations of a sea-sponge inspired skeletal lattice, its optimized design, and squared grid with no bracing, besides the respective stress versus strain curves. Reproduced with permission of~\citep{fernandes2021mechanically}. Copyright 2020 Springer Nature. (B) High buckling modes activated simultaneously in macroscale 3D printed specimens via Polyjet. Reproduced with permission of~\citep{janbaz2019ultra} under Creative Commons Attribution 3.0. (C) 3D printed lattice shoe sole application. Reproduced with permission of~\citep{weeger2019digital}. Copyright 2018 Elsevier.}
	\label{fig:05}
\end{figure*}
Moreover, fractal architected materials from one to three orders of cuts, as shown in Fig.~\ref{fig:04}(D), grasped around 88\% of shape recoverability. Although an increased fractal order has produced a decrement in bending stiffness, their structural integrity was not affected~\citep{zhang2021harnessing}. These cutting-edge configurations also guaranteed energy dissipation properties. Heterogeneous architected configurations, made of the same rigid-base material, showed enhanced capabilities with respect to those with a uniform unit cell layout. The strength from a certain type of lattice can act together with the constant post-yielding stress plateau given by another one. For example, the combination of face-centered cubic (FCC) lattices with body-centered cubic (BCC) cells~\citep{alberdi2020multi}, or with dual-rhombic (RD) topologies~\citep{white2021interpenetrating}. This latter pair (FCC+RD) formed the so-called interpenetrating lattices (IPL), with energy absorbed six times higher than the achievable values by isolated FCC or RD structures. IPL enabled a load redistribution before a sudden collapse because each former lattice configuration experienced failure at different strains~\citep{white2021interpenetrating}, as shown in Fig.~\ref{fig:04}(E) and (F). 

\begin{figure*}[!h]
	\centering
	\includegraphics[width=1\textwidth]{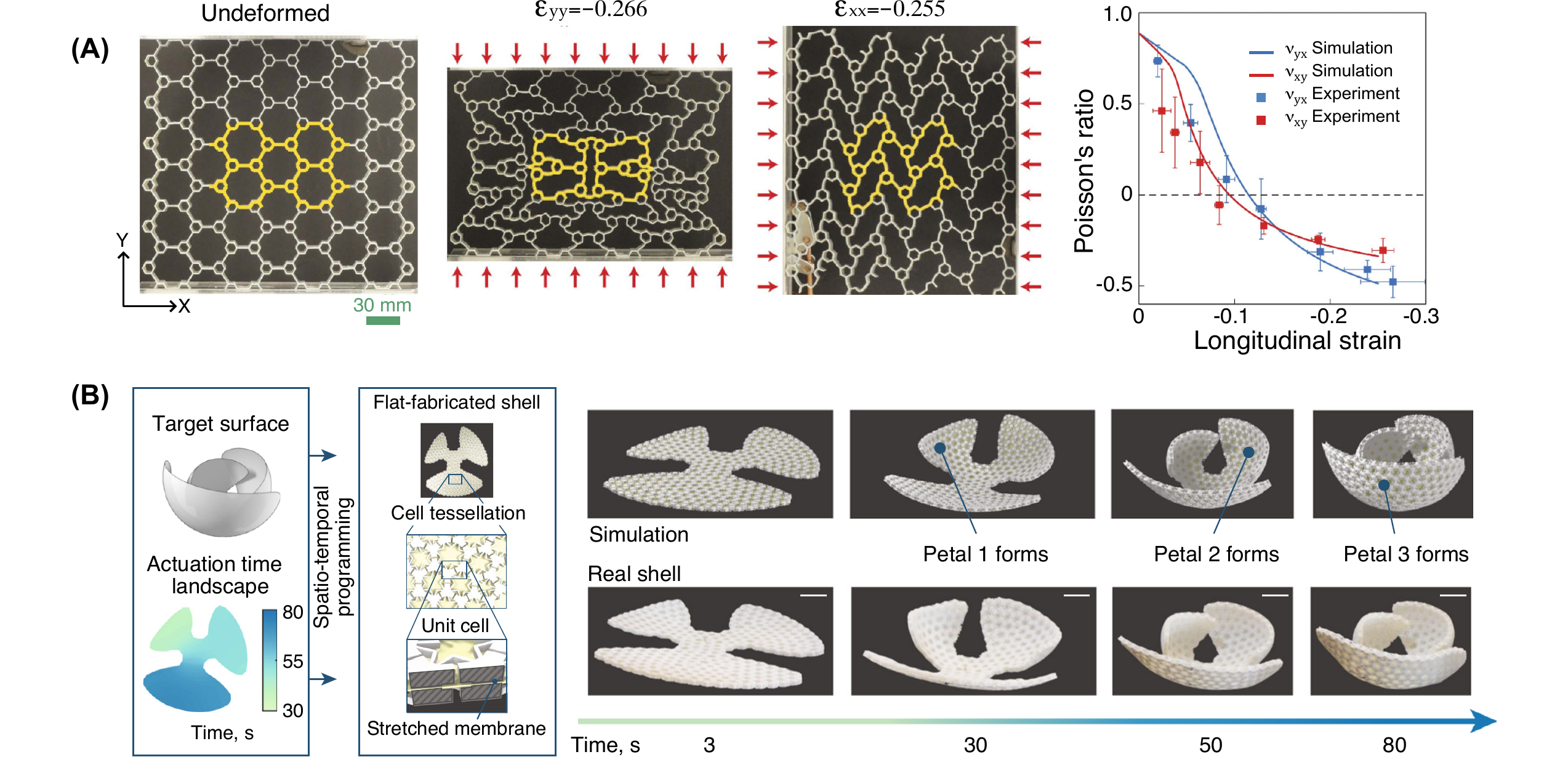}
	\caption{(A) Hierarchical honeycomb and exhibited NPR ranges. Reproduced with permission of \citep{mousanezhad2015hierarchical} under Creative Commons Attribution 4.0 International License. (B) Initial morphing shell with its internal configuration and the shape transformation process over time. Reproduced with permission of~\citep{guseinov2020programming} under Creative Commons Attribution 4.0.}
\label{fig:06}
\end{figure*}

\subsubsection{Soft lattices and instability control}

Flexible photopolymers have led to the fabrication of material structures able to withstand larger deformations, due to expected elongation at break values around 200\% (see Table~\ref{Table:3}). Soft lattices with variations in geometrical parameters (e.g. smaller ligaments diameters and larger spacing) can obtain specific elasticity profiles, for instance, the stiffness and Young’s modulus of human tissues~\citep{johnson20183d}. 
Besides the expected bending or stretching responses, different types of flexible architected structures have shown hyperelastic behavior, dominated by buckling and postbuckling regimes~\citep{jamshidian2020multiscale}. This fact enabled more buckling resistant structures without increasing the material volume along the former ligaments (e.g. sea-sponge inspired grids in the range of strains around 6\%)~\citep{fernandes2021mechanically}, as presented in Fig.~\ref{fig:05}(A). In addition to the geometrical features influence, the incorporation of instabilities allows novel capabilities, such as reversible transformations and geometric nonlinearities under specific loading conditions~\citep{valdevit2018architected}, even elastic waves control or tunable energy absorption structures~\citep{Bertoldiannurev2017}. For example, soft cellular materials,  shown in Fig.~\ref{fig:05}(B), can control buckling mechanisms by activating high levels of instability different from the initial ones and, at the same time, adjusting ranges of force actuation and amplitude without imposing extra fixtures or boundary conditions~\citep{janbaz2019ultra}. Furthermore, new nonlinear approaches have been proposed to assess performances under large deformations, for practical applications (e.g. digital material lattices for shoe soles, as displayed in Fig.~\ref{fig:05}(C))~\citep{weeger2019digital}. 

\subsubsection{Auxetic and shape programmable lattices}
During the last years, auxeticity has inspired the design of responsive lattice configurations with particular and counter-intuitive stretching or compressive deformation patterns under loading action. Such artificially-made systems are characterized by a negative Poisson's ratio \textit{(NPR)}~\citep{papadopoulou2017auxetic,ashjari2017auxetic}.
Diverse manufactured auxetic lattice structures, such as convex-concave foams ~\citep{cui2018mechanical}, hierarchical honeycombs~\citep{mousanezhad2015hierarchical}, or  unit cells with sinusoidal ligaments~\citep{jiang20203d,li2017harnessing,chen2017lattice}, have exhibited tailorable \textit{NPR} over large strains (approx.$-0.1\leq NPR\leq-0.7$).
Sinusoidal-shaped ligaments (characterized by a wave amplitude $A$, span $L$, span ratio $A/L$, wavelength number $k$, and bending beam stiffness $B$) provided tunable mechanical properties to lattice structures. Modifications in these geometrical parameters allowed tunable values of NPR and the transition between stretch and bending-dominated behavior~\citep{cui2018mechanical,jiang20203d,li2017harnessing,chen2017lattice}. 
Similarly, higher levels of hierarchy contributed to the Poisson’s ratio reduction and the consequent auxeticity~\citep{mousanezhad2015hierarchical}, as displayed in Fig~\ref{fig:06}(A).
Complex active lattice structures have been envisioned based on the thermo-viscoelastic responsive nature of photopolymers. Rectangular and triangular horseshoes patterns~\citep{wang2020design}, anti-tetrachiral, re-entrant, and curled honeycombs ~\citep{wagner2017large,dong2021design}, have developed shape recoverability. It has been observed that once NPR and certain large deformations are achieved, these structures can return to their original configuration after being subjected to an external stimulus (e.g. heat). Phase evolution models were considered to describe the shape changes produced by geometrical and thermomechanical non-linearities.
Moreover, the morphing process can also be controlled by data-driven approaches, with initial input parameters, such as a final target configuration, expected deformation rates (from experimental data), and specific actuation time. Thereby, once previously stretched shells were in contact with a heat source, a transformation from a flat into a closed-curved shape was observed for approximately 80 seconds~\citep{guseinov2020programming}, as described in Fig.~\ref{fig:06}(B).  
\begin{figure*}[h!]
	\centering
	\includegraphics[width=1\textwidth]{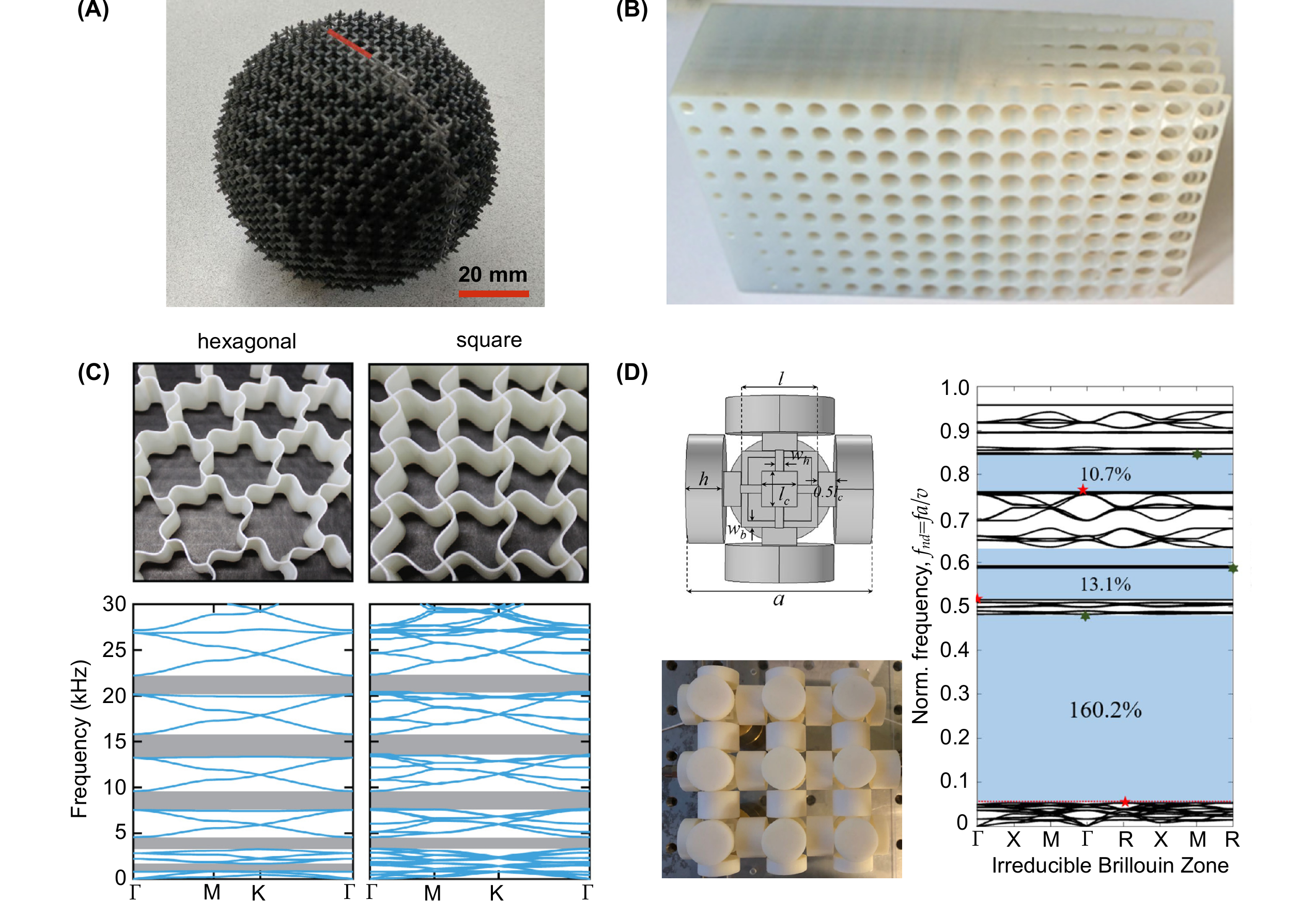}
	\caption{(A) Luneburg lenses sample for 40~kHz airborne ultrasound metamaterial. Reproduced with permission of~\citep{xie2018acoustic} under Creative Commons Attribution 4.0 International License. (B) Beam steering dielectric lens prototype. Reproduced with permission of~\citep{yi20163d}. Copyright IOP Publishing. (C) Curved ligaments lattice topologies samples and their corresponding dispersion spectra, where the zones in grey represent the obtained band gaps. Reproduced with permission of~\citep{chen2017broadband}. Copyright 2017 Elsevier. (D) Unit cell scheme (square cores) and lattice sample, with the respective band spectra and reached band gaps. Reproduced with~\citep{lim2021phononic} permission under Creative Commons Attribution 4.0 International License }
	\label{fig:07}
\end{figure*}

\subsubsection{Wave propagation control lattices}
MJ has been applied for prototyping acoustics, optics, or vibration control architected devices for later experimental validation. For example, ultrasound lenses for 40~kHz airborne ultrasound formed by 3D cross-shaped lattices~\citep{xie2018acoustic} (as shown in Fig.~\ref{fig:07} (A), chiral channels in acoustic metamaterials~\citep{peri2019axial}, thin-walled cellular structures with high sound absorption (300 - 600~Hz range)~\citep{vdovin2017implementation}, two types of all-dielectric lenses to modify wave propagation in a wide frequency range (8 to 12~GHz), beam steering and conformal arrays \citep{yi20163d}, as seen in Fig.~\ref{fig:07}(B).
Alternatives to phononic crystals with omnidirectional band gaps have been explored, such as tunable sinusoidal lattices with multiple band gaps~\citep{chen2017broadband,chen2017lattice} or vibration control metastructures with an improved bandwidth up to 160\%~\citep{lim2021phononic}, as illustrated in Fig.~\ref{fig:07}(C) and (D), respectively.   

\subsection{Multi-material 3D printed lattices}

An outstanding capability of MJ is the manufacturing of multi-material parts during one single printing stage~\citep{Stratasys1}. Nowadays, this procedure has been extended to lattice fabrication. It is important to remark that the strong interface between different photopolymers is produced by the introduction of particles from the secondary material into the base one. Therefore, the resultant mechanical properties are produced in smooth transitions~\citep{mueller2017mechanical}. Thus, there is a growing number of multi-functional architected material structures with enhanced performances, as it is mentioned in the following subsections.

\subsubsection{Reinforced composites with rigid lattices}
\begin{figure*}[h!]
	\centering
	\includegraphics[width=1\textwidth]{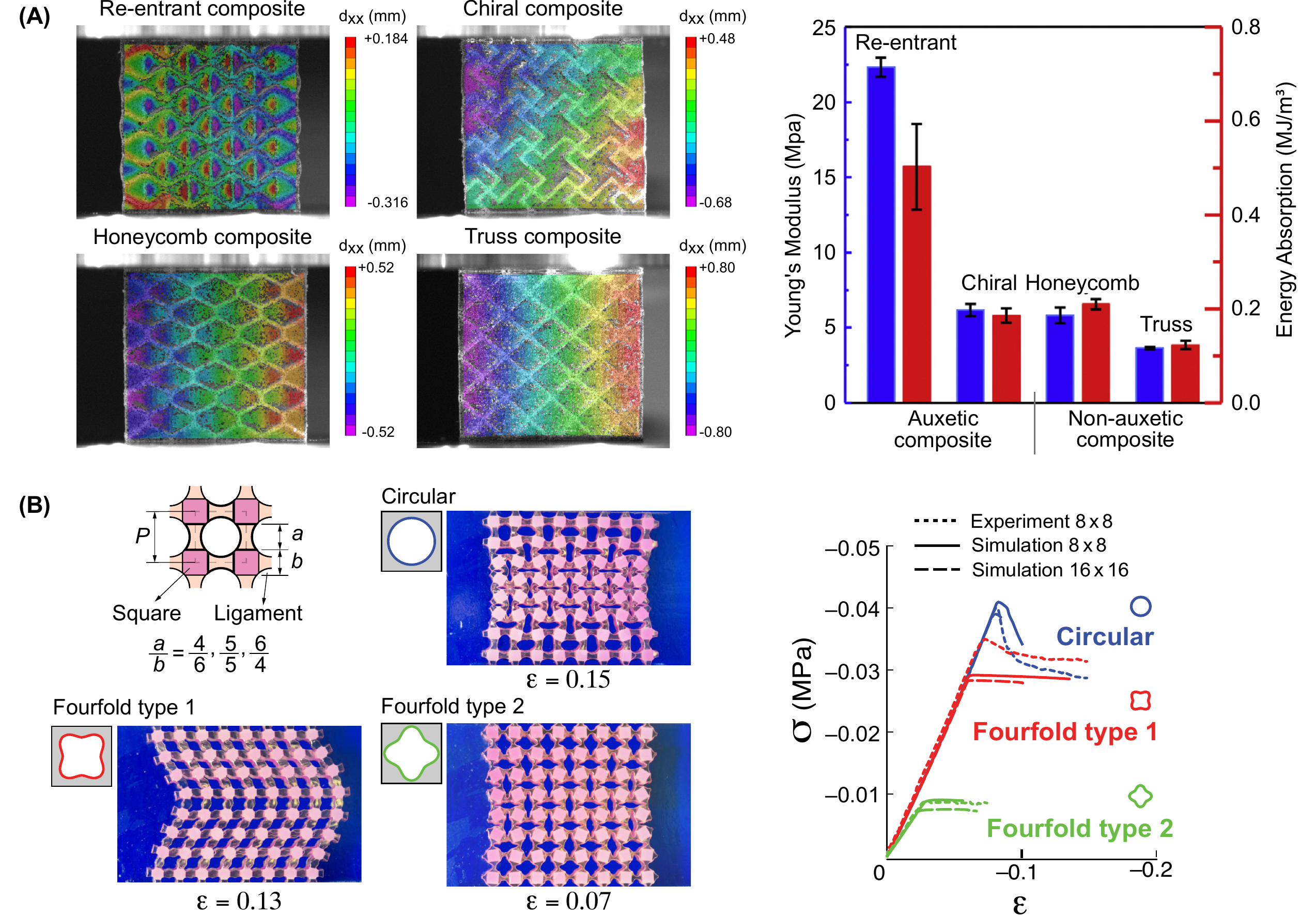}
	\caption{(A) Displacement contours at $\epsilon=0.05$ of reinforced composites (made of a soft matrix reinforced by rigid auxetic and non-auxetic lattices) displayed by digital image correlation and their respective mechanical properties comparison. Reproduced with permission of~\citep{li2018exploiting}. Copyright 2018 Elsevier. (B) Unit cell multi-material configuration, and examples of instability modes with soft ligaments in circular and fourfold tessellations, plus their obtained stress-strain curves. Reproduced with permission of~\citep{janbaz2018multimaterial}. Copyright 2018 American Physical Society.}
	\label{fig:08}
\end{figure*}

The concept of reinforced soft phases with different stiff material fractions has been developed, inspired by a variety of novel composite designs fabricated with MJ~\citep{gu2016biomimetic,zaheri2018revealing,jia20193d,yan2021recent,li2018enhanced}.
Rubbery-compliant matrices have been reinforced by several rigid polymeric lattices. In particular, TPMS with IWP-P surface type (as interpenetrating phases)~\mbox{\citep{al2017mechanical,dalaq2016mechanical}}, chiral and re-entrant auxetic honeycombs~\citep{li2018exploiting} (shown in Fig.~\ref{fig:08}(A)), Hexaround and Warmuth~\citep{albertini2021experimental}. The combination of soft infills into glassy lattices leads to more delayed densification with respect to the unreinforced counterparts. As a consequence, higher energy absorption efficiency can be expected and, at the same time, improved mechanical properties, such as increased Young’s modulus or peak stresses. A very resistant bonding between the multi-material elements was observed~\citep{dalaq2016mechanical}.

\subsubsection{Lattices with different stiffness components}
The combination of stiff and flexible elements opens alternatives to obtain lattices with improved and tunable mechanical properties. Besides the variations in geometrical parameters, the distribution of rigid and rubbery phases along lattices members allowed the tunability of elastic modulus and Poisson's ratio~\citep{mirzaali2018multi,jiang20183d,bossart2021oligomodal}, or the control of instabilities and buckling modes~\citep{janbaz2018multimaterial}, as seen in Fig.~\ref{fig:08}(B). Indeed, the increment of hard phases with respect to soft phases tended to increase the loading capacity.
The combination of stiff material members and compliant hinges allowed constant zero-energy modes attainable without the use of external actuators. In particular, two selective actuation modes were obtained by controlling the viscoelastic properties of the joints, where their stiffness increased with the strain rate. This fact is beneficial to controlling Poisson's ratio in a certain strain range ($0.3\leq \epsilon \leq0.45$). Hence, either negative or positive values were observed at slow and fast compression rates, respectively~\citep{bossart2021oligomodal}. Furthermore, Hoberman sphere$^\text{\textregistered}$~\citep{hoberman1990reversibly}, inspired structures achieved multiple and tunable broadband gaps. The inclusion of rigid or soft intermediate rods allowed variable stiffness that can adjust the size of the band gaps~\citep{li2019phononic}. In addition, similar multi-material unit cells may exhibit a variable coefficient of thermal expansion (CTE), which depends on the differences in geometry, Young's modulus, and CTE among the former lattice elements~\citep{li2018hoberman}.

\begin{figure*}[h!]
	\centering
	\includegraphics[width=1\textwidth]{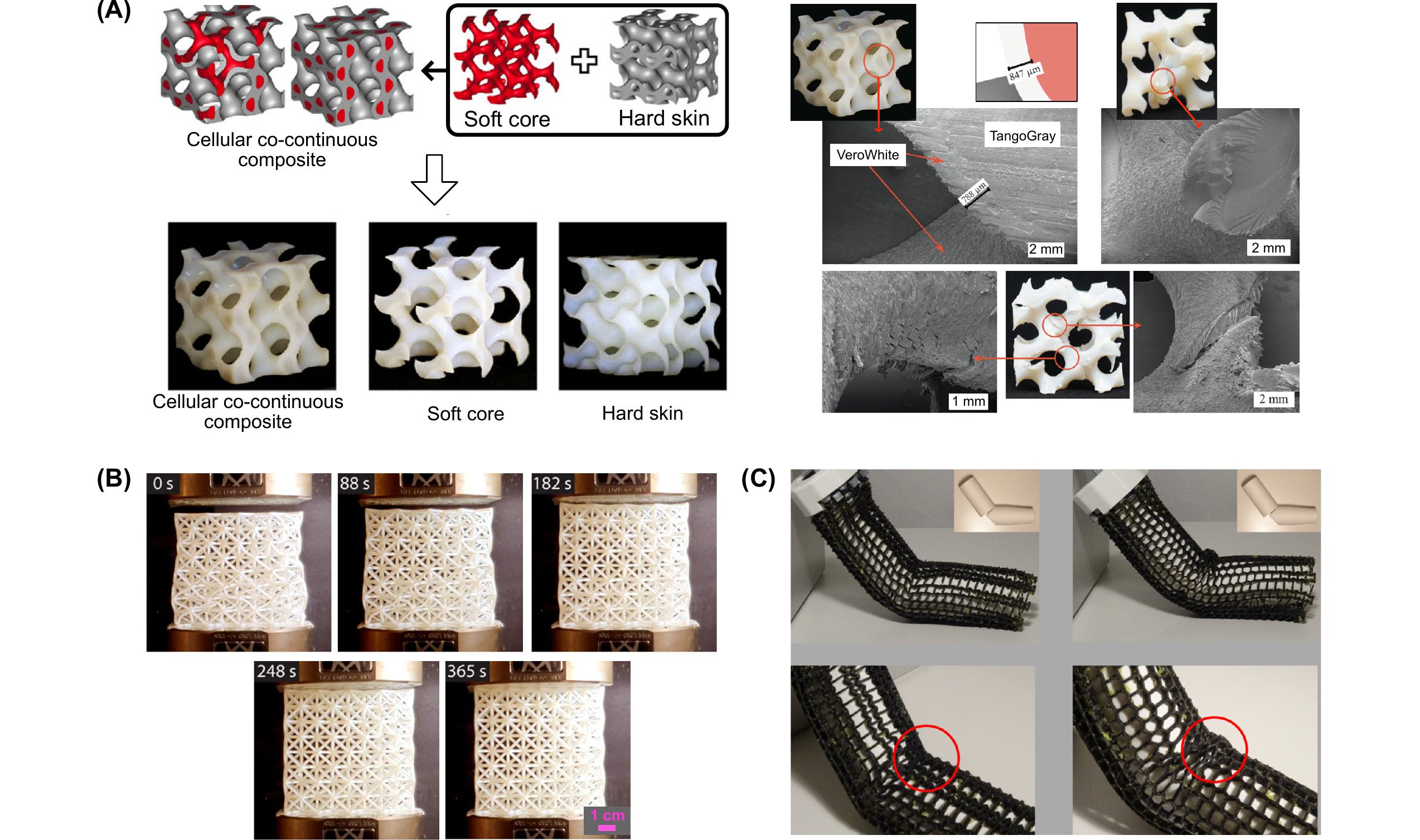}
	\caption{(A) Configuration of cellular graded TPMS made of bonded soft and hard material substructures. Reproduced with permission of~\citep{al2018nature}. Copyright 2017 WILEY-VCH Verlag GmbH. (B) Morphing shape recovery of a graded structure due to increasing temperature. Reproduced with permission of~\citep{lumpe2021computational} under Creative Commons Attribution 4.0 International License. (C) 3D printed graded distribution of radial unit cells combining negative and positive Poisson's ratios applied in a hinged arm exo-suit (left part) and compared to a hexagonal pattern (right part). Reproduced with permission of~\citep{hedayati20213d}. Copyright AIP Publishing. }
	\label{fig:09}
\end{figure*}
\subsubsection{Functionally graded and active lattices}
Non-affine deformations in mechanical metamaterials are produced by increments of harder material fractions with respect to soft ones. Thereby, higher values of elastic modulus and stress concentrations were produced. On the other hand, properties like auxeticity are enabled~\citep{mirzaali2020non}.
For this reason, functionally graded materials or corner-free structures represent an option to control the effects of high localized stress distributions to preserve structural integrity while flexibility is not limited~\citep{mirzaali2020non}. For instance, Schoen’s gyroid TPMS graded unit cells with a variable hard shell and a soft core, as shown in Fig.~\ref{fig:09} (A), followed this criterion~\citep{al2018nature}. Moreover, cylindrical optimized struts composed of a flexible outer cover and rigid-inner cores were extended to Voronoi lattices. In this case, the energy absorption capability was improved up to 38 times with respect to single material strut lattices, but prevented the reduction of stiffness and strength~\citep{mueller2018stepwise}.  
The concept of dual and/or graded material auxetic lattices has also been exploited~\citep{wang2015designable,saxena20173d,hedayati20213d}. The equivalent Young’s modulus and NPR can be modified by varying the stiff material gradient content along the lattice members according to the loading condition (e.g. cells with flexible joints and rigid ligaments). 
Furthermore, 3D-printed active architectures can be fabricated by the combination of materials with diverse responses under external stimuli. Variations in stiffness due to high temperatures allowed the design of morphing graded structures programmed to reach a target state~\citep{lumpe2021computational}, as displayed in Fig.~\ref{fig:09}(B). Similarly, anti-chiral lattices with a tailorable coefficient of thermal expansion (CTE) were achieved by tuning the lattice geometry together with the use of photopolymers with different CTE~\citep{wu2016isotropic}. Thermal expansion, as well as auxetic properties, are not scale-dependent. Thus, they can be extended from macro to nanoscales~\citep{wu2016isotropic}. 
The development of practical applications inspired by graded lattices has been applied in garments. For example, the concept of a protective hinged arm was developed by radial graded auxetic patterns (RNP), that combine negative and positive Poisson's ratio~\citep{hedayati20213d}. They prevented wrinkles and lateral displacements when the elbow is bent (Fig.~\ref{fig:09}(C)-left), contrary to what is observed in other patterns (e.g. hexagonal as seen in Fig.~\ref{fig:09}(C)-right).

\section{Challenges, advantages, and limitations}\label{chall}
MJ allows the use of photopolymers with different mechanical properties, and it has paved the way for the fabrication of architected lattice materials with a variety of performances, from high-strength to morphing shape structures as summarized in Appendix~\ref{summary:AL}, Table~\ref{tab:SM}. An integrated overview of the main challenges, future works, and limitations that MJ lattices manufacturing faces are exposed in the subsequent sections. Firstly, in terms of the printing, modeling, and experimental processes, then, outlining the actual results and potential for future applications. 

\subsection{MJ architected lattices: Printing process and experimental work challenges }

Defects, imperfections, and anisotropy on 3D printed lattices are induced by the AM process. Then, the numerical model predictions tend to differ from the mechanical performances exhibited by the final object. In particular, the assessment of Young's modulus or tensile strength. There are few lattices analyzed with models that include MJ fabrication constraints, like printing orientation effects in a transversely isotropic hyperelastic model~\citep{zhang2016transversely}. However, it is suggested the extension of this approach towards shear and/or biaxial experiments besides uniaxial tests outcomes. The influence of the UV exposure time and intensity, or post-curing stages, are crucial aspects to be also considered and deeply analyzed. Thereby, a holistic design approach can be oriented towards a more realistic description of the lattice's mechanical properties.
The printing quality and dimensional accuracy of the lattice elements are very sensitive to the cleaning state of the printer components. Maintenance tasks are mandatory to obtain precise printed parts. Before and after operating the respective printers, the print heads must be cleaned to avoid blocked nozzles. Thus, a free resin flow, less surface roughness, and depressions on the object can be expected to cause uneven printed surfaces. Similarly, resin residues on the lamp glasses must be removed because they hinder the passing of UV light. Then, the curing process is affected, it tends to be not uniform along the deposed layers and it provokes defects in the desired lattice geometry.
The majority of existing MJ lattices have been fabricated mostly with  glassy photopolymers (e.g. VeroWhite), which have been characterized by tensile, compressive, and flexural strength tests. Although the viscoelastic behavior of resins represents a level of complexity for experiments~\citep{vu2018characterizing}, the same procedure can be extended to explore the mechanical performances of other available photopolymers (e.g. soft and rigid composites/digital materials) and emerging eco-friendly photocurable resins~\citep{PIERAU2022101517,voet2021sustainable,ChiulanPhotopolymerization}. The latter represent a potential alternative that will require material characterization studies to guarantee high performance and sustainable 3D printed parts at the same time. Moreover, it is also necessary to study more in-depth the fatigue life and aging effects of lattice members, as well as large strain deformations, failure, and non-linearities. These aspects are still challenging, especially for rubber-based components. Thus, detailed constitutive models can be found to describe more precisely the properties of 3D printed soft meta-structures and to assess accurately their service life. 
A significant number of architected lattices presented in this review have been evaluated under quasi-static loads during the experimental and numerical simulation phases. Dynamic tests like impact or wave propagation tests still remain in an area that could be further explored. 
In spite of a growing number of preliminary studies and cutting-edge works related to graded and multi-materials, the already achieved experimental results require complementary validation, either numerical or analytical. More extended assessments of interfaces of multi-material parts are indeed necessary to be integrated for design and failure considerations. Thereby, it is possible to produce new types of composites and multi-material lattices for several engineering areas with outstanding properties and more controlled damage.

\subsection{MJ architected lattices: manufacturing limitations }\label{man:limits}
There are fabrication constraints that need to be considered during the design stage of high-quality architected lattice structures. An evaluation related to the limitations in terms of geometry and cost estimation is presented as follows. 
\subsubsection{Geometrical limits}
Printers manufacturers' guidelines suggest not considering load-bearing elements with a wall thickness below 0.5~mm~\citep{3Dsystemsbestpractice} or 0.6~mm~\citep{Stratasysguideline} due to potential fragility and risk of damage during post-processing. We printed different samples with glassy photopolymers via the Polyjet technique (Stratasys J750 printer) to determine limiting thickness values for load-bearing and floating ligaments which are part of the MJ architected lattices structures. 
\begin{figure*}[h!]
	\centering
	\includegraphics[width=1\textwidth]{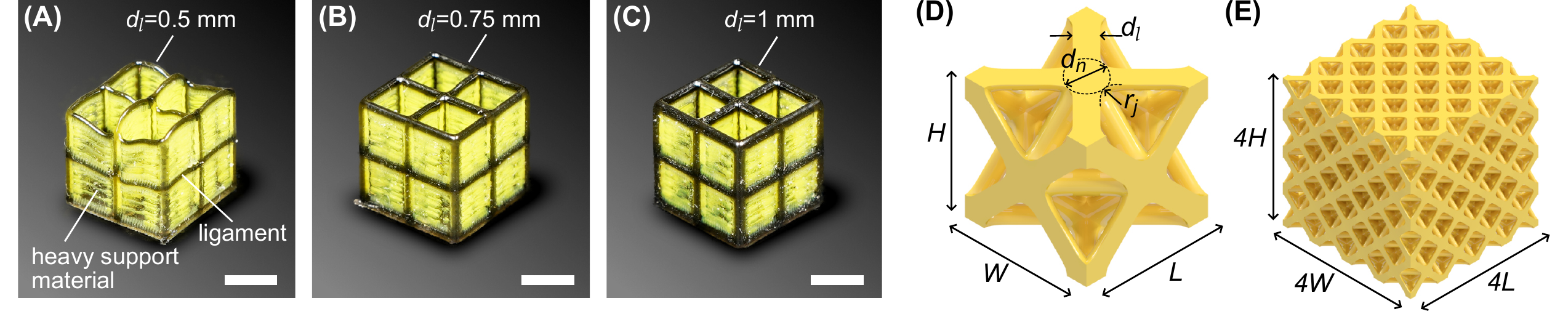}
	\caption{Thickness limits for MJ printed lattice structures. Printed cubic lattices  (10x10x10~mm) with ligaments diameters of (A) 0.5~mm, (B) 0.75~mm, and (D) 1~mm and fabricated with a heavy density grid support by a Stratasys J750. Very thin-walled floating (horizontal) ligaments with thickness less than 1 mm tend to bend during the jetting process. Scale bar: 5~mm.  (D) Octet truss unit cell and (E) the corresponding three-dimensional 4x4x4 tessellation of octet unit cells considered in the printing cost analysis. The considered geometrical parameters are $W=$30~mm, $L=30$~mm and $H=30$~mm, ligaments diameter $d_l=$4.0~mm, node diameter $d_n=$8.0~mm, and node-ligament junction radius $r_j=$3.1~mm.}
	\label{fig:09b}
\end{figure*}
Cubic lattices (10x10x10~mm) were fabricated with ligament diameters of 0.5~mm, 0.75~mm, and 1~mm. The selected base materials and support material were respectively VeroBlack (rigid photopolymers) and SUP706B, and different levels of grid density support were considered (lite, standard, and heavy). The printed structures with a heavy density grid support are shown in Fig.~\ref{fig:09b}(A), (B) and (C) for ligaments diameters of 0.5~mm, 0.75~mm, and 1~mm, respectively. In spite of using the strongest support grid (heavy level), very thin-walled floating (horizontal) ligaments with a thickness of less than 1~mm tended to bend during the jetting process since the amount of material is not enough to form a strong base substrate for the upcoming layer, and the created bond is weak. Then, the resultant geometry appears deformed during the hardening stage with the UV light curing. Thus, the resulting printed ligaments present defects. It was also observed that ligaments with a thickness of 0.5~mm became fragile and broke easily during the support removal operations. Furthermore, we have tested axially the printed grids. In the case of ligaments with diameters of 0.5~mm and 0.75~mm, the imperfections attributed to the mentioned fabrication defects contributed to a faster ligaments' failure.  
Therefore, diameters of 1~mm provide a safe margin to preserve the MJ lattice geometry and prevent defects, especially for horizontally arranged ligaments, as shown in Fig.~\ref{fig:09b} (C).
\begin{table*}[!b]
 \centering
 	\caption{Total printing time, material, energy  and cost estimation of an octet truss unit cell (30x30x30~mm, ligaments diameter of 4.0~mm), and corresponding three-dimensional 4x4x4 tessellation fabricated both via MJ and FDM. All costs were calculated in USD(\$) without VAT. A proposed $L_{cost}=15$~\$ value can be added to Total cost ($TC$) for labor cost inclusion. See Appendix~\ref{app_MJ_cost} for calculation details.}
 {\renewcommand{\arraystretch}{1.5}
 \begin{tabular}{*1{>{\centering\arraybackslash}m{0.02\textwidth}}*1{>{\centering\arraybackslash}m{0.02\textwidth}}*1{>{\centering\arraybackslash}m{0.19\textwidth}}*9{>{\centering\arraybackslash}m{0.055\textwidth}}}
\otoprule
& \multicolumn{2}{c}{Printing mode} & Printing time (hours) & Model Material (g) & Support Material (g) & Material cost (\$)& Energy (MJ)& Energy cost (\$) & Indirect cost (\$) & Total Cost (\$)& Total Cost (\$/cm$^{3}$)\\
 \hline
\parbox[t]{2mm}{\multirow{5}{*}{\rotatebox[origin=c]{90}{\textbf{Octet cell}}}} &
\parbox[t]{2mm}{\multirow{4}{*}{\rotatebox[origin=c]{00}{\textbf{MJ}}}}
&  High-Mix Glossy       & 2.02    & 54   & 33   & 20.93    & 3.97   & 0.27   & 13.33   & 50.44  & 9.46\\
& &High-Mix Matte        & 2.02    & 57   & 48   & 23.60    & 3.97   & 0.27   & 13.33   & 55.13  & 10.34\\
& &High-Quality Glossy   & 3.62    & 95   & 45   & 35.33    & 7.04   & 0.48   & 23.90   & 86.57  & 16.23\\
& &High-Quality Matte    & 3.62    & 98   & 59   & 37.88    & 7.04   & 0.48   & 23.90   & 91.06  & 17.08\\
\cline{2-12}
& \parbox[t]{2mm}{\multirow{2}{*}{\rotatebox[origin=c]{00}{\textbf{FDM}}}}  
&   0.05~mm UD layers   & 5.88    & 6.70  & 5.53   & 1.22    & 2.12   & 0.15   & 0.15   & 1.51   & 0.28\\
& & 0.15~mm layers      & 2.30    & 6.70  & 5.17   & 1.16    & 0.83   & 0.06   & 0.06   & 1.28   & 0.24\\
 \otoprule
 \parbox[t]{2mm}{\multirow{5}{*}{\rotatebox[origin=c]{90}{\textbf{Octet lattice}}}} &
\parbox[t]{2mm}{\multirow{4}{*}{\rotatebox[origin=c]{00}{\textbf{MJ}}}} 
&  High-Mix Glossy      & 25.92    & 1372    & 1542   & 612.21    & 49.84   & 3.43   & 171.28   &  1252.19   &  3.67\\
& &High-Mix Matte       & 27.77    & 1518    & 2005   & 711.51    & 53.39   & 3.67   & 183.51   &  1439.44   &  4.22\\
& &High-Quality Glossy  & 45.32    & 1795    & 1642   & 758.06    & 87.07   & 5.99   & 299.50   &  1639.66  &  4.80\\
& &High-Quality Matte   & 48.65    & 1961    & 2097   & 862.80    & 93.47   & 6.43   & 321.53   &  1846.49  &  5.41\\
\cline{2-12}
& \parbox[t]{2mm}{\multirow{2}{*}{\rotatebox[origin=c]{00}{\textbf{FDM}}}}  
& 0.05~mm UD layers   & 290.93    & 424.87   & 299.88   & 69.30    & 104.74   & 7.20   & 7.28   & 83.78   & 0.25\\
& & 0.15~mm layers    & 123.12    & 421.31   & 285.21   & 66.79    & 44.32    & 3.05   & 3.08   & 72.92   & 0.21\\
 \hline
 \end{tabular}}
 \label{Table:2}
 \end{table*}
\subsubsection{Cost analysis}
The high fabrication cost is an aspect that has limited the use of MJ in the AM market with respect to other techniques. This fact has also been extended to lattice manufacturing. There are additional factors that contribute to the increment of the final cost of a MJ printed part, besides the raw materials and energy consumption price. Therefore, the Total Cost ($TC$) estimation of 3D printed lattices via MJ, also includes direct and indirect costs based on the printing time involved in their fabrication, and it can be calculated by the formula \mbox{$TC=I_{cost} + L_{cost}+\omega\cdot (M\cdot M_{cost})+(E_{total}\cdot E_{cost})$}~\citep{baumers2015modeling}. Where $I_{cost}$ refers to indirect costs (e.g. machine depreciation), $L_{cost}$ is the approximated labor cost, \textit{M} is the amount of used material with the respective cost $M_{cost}$, $\omega$ is a factor that considers the wasted material accumulated on the roller after each pass to uniform the deposited layers~\citep{baumers2015modeling}, $E_{total}$ is the amount of energy consumed during the process with its price $E_{cost}$. The target geometries for this case study were (i) an octet truss unit cell and (ii) the corresponding three-dimensional 4x4x4 tessellation of octet unit cells.
With reference to the geometrical parameters depicted in Fig.\ref{fig:09b} (D) and (E), the dimensions of octet unit cell are $W=$30~mm, $L=30$~mm and $H=30$~mm, ligaments diameter $d_l=$4.0~mm, node diameter $d_n=$8.0~mm, and node-ligament junction radius $r_j=$3.1~mm.
The cost estimation and printing time are presented in Table~\ref{Table:2}. See further details of the cost estimation in Appendix~\ref{app_MJ_cost}. The employed materials for the MJ printing were VeroYellow RGD-836 and SUP-706B for model and supports, respectively. The material and energy costs assessment were done by following Italy market quotations for late 2021 and energy prices for March 2022, respectively. The printing modes considered were High-Mix (layers height=27 microns) and High-Quality (layers height=14 microns) in order to assess the lattice fabrication with the highest resolution possible. In addition, the two available surface finishing options, Glossy and Matte, were applied.  Moreover, we provide an estimated cost and printing time of the same samples if they were fabricated by more widely used and affordable techniques in the AM market, such as FDM. Thereby, it was possible to establish a comparison among the printing costs obtained by using a Stratasys J750 Polyjet printer, with those expected by printing the same structures with a cheaper FDM 3D printer Prusa i3 MK3S. For this reason, the imposed dimensions of the case study geometries were selected based on the feasibility of manufacturing in both techniques. For FDM processes is not possible to achieve a satisfactory resolution in detailed structures with small dimensions, like octet cells. In the case of FDM cost assessment, the employed materials were Prusament PC Blend Jet Black (filament with diameter=1.75~mm) and soluble supports Verbatim BVOH, without considering a waste factor $\omega$.  The printing modes were Ultra-detail (UD) and Quality with 0.05~mm and 0.15~mm layers height, respectively.
    Additional post-processing costs (e.g. cleaning devices, alkaline solutions) were not considered in both printing processes. The purpose was to obtain specifically raw costs attributed to the model and support materials plus the machine, in terms of energy consumed and depreciation in time. Similarly, the labor cost, $L_{cost}$, was not included in the Total Cost ($TC$) estimation, shown in Table~\ref{Table:2}. It is a subjective parameter that remains constant, either for the fabrication of one unit cell or the corresponding tessellation, or the printing technique. Thus, if a further estimation is required, a $L_{cost}$ value can be added to $TC$, based on the salary per hour of a technician (e.g. considering $S_{hour}=15~\$/$hour and an approximated time fraction for printing operations $t_{op}=1$~hour, as indicated in Appendix~\ref{app_MJ_cost}). Furthermore, the difference in total costs between MJ and FDM processes is very significant. For example, considering the 4x4x4 octet lattice case, where the costs are MJ= 3.67~\$/cm$^3$ (High-Mix Glossy) versus FDM= 0.25~\$/cm$^3$ (UD layers). This fact correlates to the expected printing resolution in terms of layer thickness (MJ=14-27~microns versus FDM=0.05-0.15~mm), and shorter fabrication time, which almost reached differences from six to ten times (MJ=25.92--48.65~hours versus FDM=123.12--290.93~hours). MJ results more cost effective by printing objects in series as the lattice case, than single isolated parts with a difference of almost three times. For instance, considering High-Mix Glossy printing mode, the estimated cost is about 3.67~\$/cm$^3$ for the 4x4x4 octet lattice versus 9.46~\$/cm$^3$ for the octet unit cell, as can be seen in Table~\ref{Table:2}. Similarly, a further cost reduction in MJ is observed when there is printing at full capacity. For instance, a 16x12x4 octet  lattice can occupy almost the 100\% of the build-tray of a Stratasys J750 MJ printer (490x390~mm). In this case, the estimated Total Cost ($TC$), considering High-Mix and High-Quality modes, is between 2.60-2.87~\$/cm$^3$, which is almost 30-40\% less than the observed cost of the smaller 4x4x4 octet lattice, about 3.67-4.80~\$/cm$^3$. As more objects are placed along the longitudinal direction of the tray, the fastest they are fabricated during a single horizontal pass of the print head. Thus, the total cost can be optimized. 

 \begin{figure*}[!h]
	\centering
	\includegraphics[width=1\textwidth]{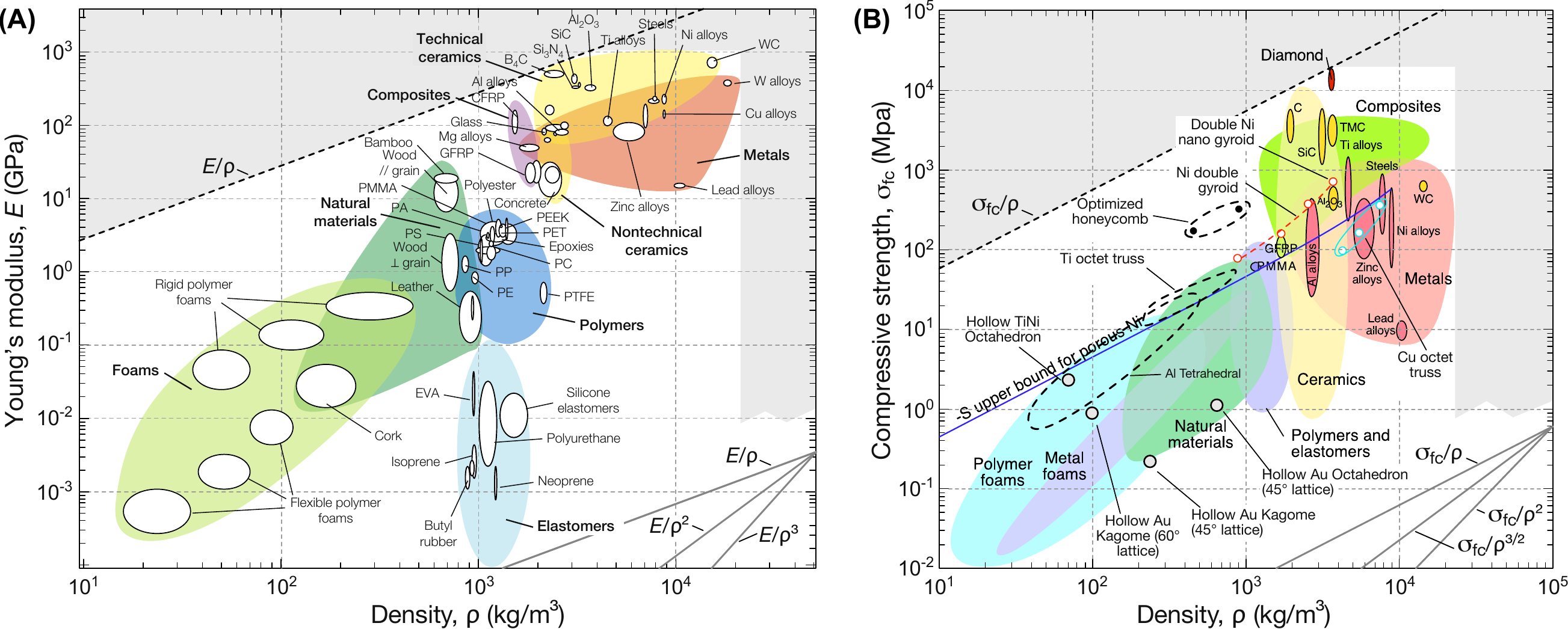}
	\caption{(A) Materials development and unattainable space regions are plotted in Young’s modulus vs. density graph. (Plot created by using CES EduPack 2019, ANSYS Granta~\copyright 2020 Granta Design). (B) Recent strongest/lightest materials strength versus density space curves, from research near to nano/ micro lattices. Adapted from~\citep{fleckdeshpandewebinarEML,khaderi2017indentation}. The upper theoretical limits reported in sub-figures are defined, respectively, by the lines with a slope of $E/\rho$ and $\sigma_{fc}/\rho$~\citep{ashbymaterials}, being tangent to diamond which is traditionally considered the stiffest and strongest material.}
	\label{fig:10}
\end{figure*}

\subsection{MJ architected lattices: outcomes and potential}

There is an increasing number of research works related to architected lattice materials, starting from the improvements of traditional cellular materials towards the exploration of optimization approaches. However, there is a permanent target to obtain high ductility by decreasing the length scale of the lattice, thereby filling existing gaps delimited in the Ashby charts~\citep{fleckdeshpandewebinarEML,khaderi2017indentation}, as shown in Fig.~\ref{fig:10}(A) and (B). The upper theoretical limits reported in Fig.~\ref{fig:10}(A) and (B) are defined, respectively, by the lines with a slope of $E/\rho$ and $\sigma_{fc}/\rho$~\citep{ashbymaterials}, being tangent to diamond which is traditionally considered the stiffest and strongest material. Current AM technologies aim to fulfill this objective by fabricating structures at meso, micro, and even nanoscale or hierarchical designs with higher orders.
Due to the high printing resolution of rigid and/or soft photopolymer objects, the MJ technique has contributed to an increasing number of tunable, high-resistant, and lightweight structures, as mentioned in this review. It also includes innovative hierarchical materials~\citep{mousanezhad2015hierarchical,chen20183d} and applications to real-life products, such as shoe soles~\citep{weeger2019digital} or exo-suits~\citep{hedayati20213d}.
For the assessment of the lattice mechanical properties, data obtained from experimental works are inserted in Ashby plots and compared to existing materials available in CES EduPack 2019, Granta Design~\citep{Granta_Design} database.

\begin{figure*}[h!]
	\centering
	\includegraphics[width=1\textwidth]{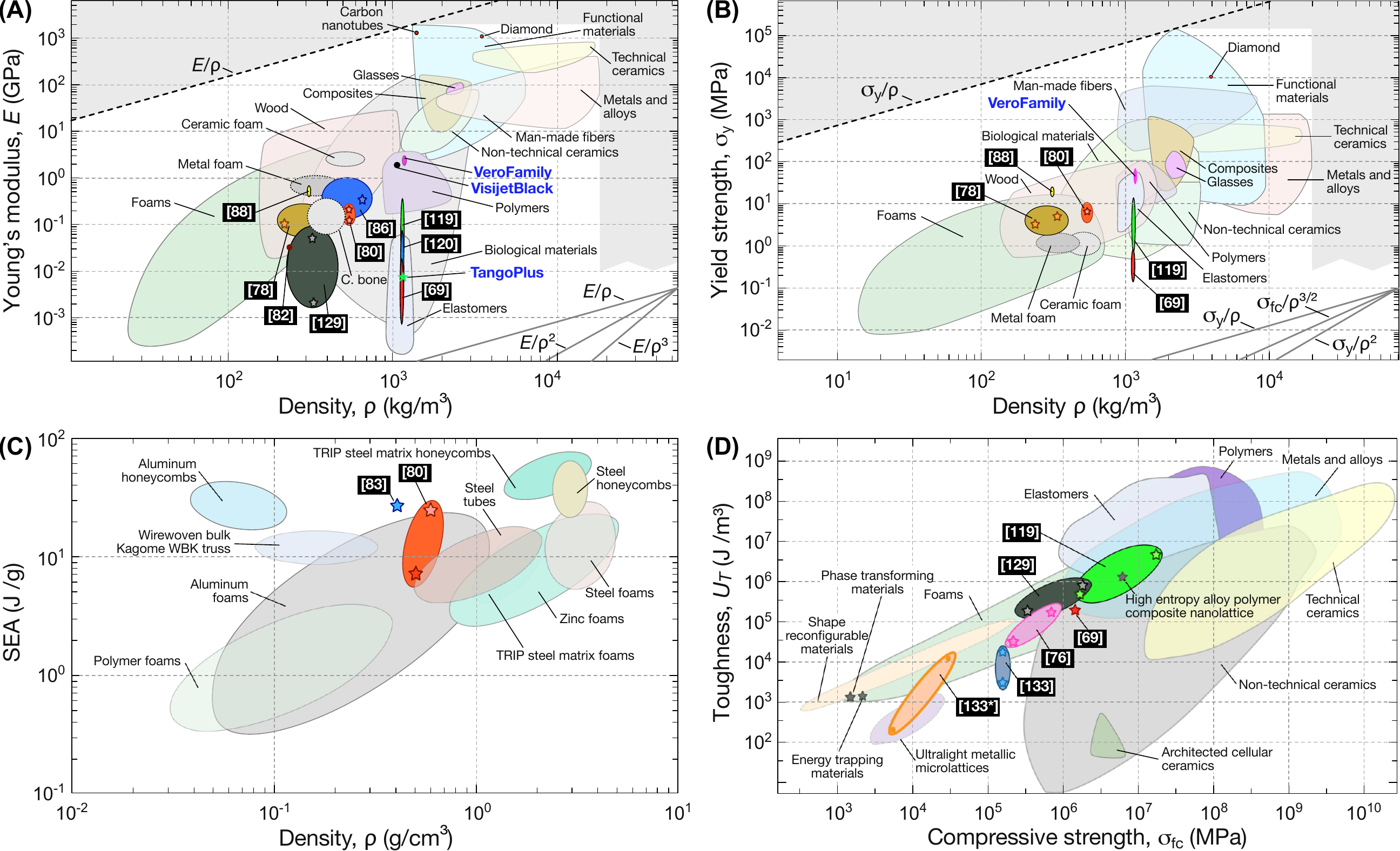}
	\caption{Ashby plots: (A) Young's modulus vs. density, (B) Yield strength vs. density, (C) Specific Energy Absorption (SEA) vs. density, and (D) Toughness vs. compressive strength. The considered MJ lattices experimental results were taken from the following works:  semi-plate~\citep{li2021additive}, TPMS with radially graded porosity ~\citep{afshar2018compressive}, TPMS~\citep{kadkhodapour2014investigating}, cubic reticulated ~\citep{alaboodi2018experimental}, IPL~\citep{white2021interpenetrating}. Multimaterial TPMS ~\citep{al2017mechanical}, and~\citep{dalaq2016mechanical}, auxetic reinforced soft composite ~\citep{li2018exploiting}, graded TPMS~\citep{al2018nature}, tailored honeycombs~\citep{kumar2019tunable} and TPMS with radially graded porosity~\citep{afshar2018compressive}, graded struts~\citep{mueller2018stepwise}, \citep{mueller2018stepwise}*(compressive strength taken up to failure), snap-fitted~\citep{liu2020maximizing}. The limit values of their envelopes are represented by stars. (A) and (B) plots were created by using CES EduPack 2019, ANSYS Granta~\copyright 2020 Granta Design. (C) plot was adapted from~\citep{eckner2020investigations}. (D) plot was adapted from~\citep{mueller2018stepwise} and~\citep{haghpanah2016multistable}. The theoretical upper limits reported in the sub-figures (A) and (B) are defined, respectively, by the lines with a slope of $E/\rho$ and $\sigma_{fy}/\rho$~\citep{ashbymaterials}, being tangent to carbon nanotubes, the existing stiffest and strongest materials.}
	\label{fig:11}
\end{figure*}
Young's Modulus or yield strength versus density plots on a log scale are inserted, as illustrated in Fig.~\ref{fig:11}. In this case, the theoretical upper limits reported in Fig.~\ref{fig:11}(A) and (B) are defined, respectively, by the lines with a slope of $E/\rho$ and $\sigma_{fy}/\rho$~\citep{ashbymaterials}, being tangent to carbon nanotubes, the existing stiffest and strongest materials. In the case of single material lattices, diverse configurations such as semi-plate~\citep{li2021additive}, TPMS~\citep{afshar2018compressive,kadkhodapour2014investigating}, and cubic reticulated ~\citep{alaboodi2018experimental}, reached
levels of stiffness close to those exhibited by metal foams and biomaterials (e.g. cancellous bone). Thus, they represent a suitable solution to fabricate scaffolds for tissue growth. While multi-material TPMS~\citep{al2017mechanical,dalaq2016mechanical}, or auxetic reinforced composites~\citep{li2018exploiting}, approached high-performing elastomers. Graded TPMS lattices~\citep{al2018nature} achieved Young's modulus close to foams at less density values, as shown in Fig.~\ref{fig:11} (A). It can also be evidenced that MJ lattices obtained more elastic properties at a lower density than their solid constituent counterparts (e.g. VeroWhite, Visijet) and non-technical ceramics groups in some cases.
Moreover, improved yield strength values were also observed in structures such as semi-plate~\citep{li2021additive}, TPMS with radially graded porosity~\citep{afshar2018compressive} and cubic reticulated~\citep{alaboodi2018experimental}, with respect to available metal and ceramic foams at the same density, as displayed in Fig.~\ref{fig:11} (B). It opens the possibility to continue the development of lattices with this 3D printing technology to enhance the mechanical performances of traditional foams, even with lighter materials. Hence, some gaps inside the observed Ashby plots could be covered. However, filling a higher number of them is still a challenge that envisions future works.
Energy absorption has been one of the most advantageous mechanical properties achieved by several MJ lattices discussed in previous sections. For example, tailored honeycombs ~\citep{kumar2019tunable} or TPMS with radially graded porosity~\citep{afshar2018compressive} achieved Specific Energy Absorption values per unit mass (SEA) up to $SEA=$30~J/g, close to the exhibited ranges by steel tubes between $SEA=$ 30~J/g - 60~J/g~\citep{mallick2012failure}.  The adapted plot from~\citep{eckner2020investigations} study in Fig.~\ref{fig:11} (C) allows the comparison among the SEA of the reviewed structures and the values achieved by some architected materials. The \textit{SEA} values of MJ lattices became higher than polymeric or metal foams and steel tubes. 
Furthermore, it is possible to highlight the energy absorbed by multimaterial lattices in an Ashby plot based on~\citep{haghpanah2016multistable} research work, as shown in Fig.~\ref{fig:11} (D). TPMS~\citep{al2017mechanical,dalaq2016mechanical}, graded structures~\citep{al2018nature,mueller2018stepwise}, and the snap-fitted lattices~\citep{liu2020maximizing} exhibited values of toughness (Energy per volume) comparable to the highest grasped by foams and some groups of ceramics. Similarly, the reported compressive strength data reached the top values of cellular foams. The capability of using stiff photopolymers together with flexible materials or as the reinforcement of a soft matrix reduces the possibility of generating brittle failures. Rubbery materials withstand compressive loads under large strains, and they improve the fragile behavior attained to the hard counterparts. 
Promising results like multiple and wide bandgaps at ultralow frequency~\citep{lim2021phononic} represent an exciting alternative for filtering or damping devices. Thus, vibration control and high-impact 3D printed lattices might start to be designed and tested. However, the wave propagation in a viscoelastic media, like MJ photopolymers, demands further analysis.
As an additional remark, shape programmable lattices open the possibility to create 4D printed materials. The presented 2D works described how to exploit the ability to morph glassy polymers under external stimuli like heat. These experimental and numerical studies can be extended to 3D topologies besides the widely analyzed 2D auxetic configurations. Tunable Poisson's ratio was another outstanding property exhibited by some reviewed works with single and graded materials. In particular, the Poisson's ratio ranges between 0 (in tension) to -0.4 (in compression) for 3D lattices made of flexible digital material~\citep{li2017harnessing}, between 1 and -0.5 for hierarchical honeycombs ~\citep{mousanezhad2015hierarchical}. In the case of tubular structures a Poisson's ratio of -0.7 was observed~\citep{jiang20203d}. Furthermore, negative Poisson's ratios up to -0.8 at strains of 0.4 were achieved by the multimaterial chiral lattice (Soft-Digital Material and rigid photopolymeric elements) ~\citep{jiang20183d}. These improved mechanical properties lead to more alternatives to study in-depth the potential use of multi and digital materials, besides the commonly used glassy resins.

\section{Concluding remarks} \label{conclusions}
The present review was divided into two main parts. First, the most relevant features concerning MJ technology were introduced.  Moreover, the influence of the printing process parameters on the mechanical properties  of the 3D printed lattices has been intensely discussed based on available experimental works. In particular, the differences in tensile strength due to the anisotropy caused by the print direction. Second, an overview of the latest and more extensively studied 3D printed architected lattice structures with MJ is exposed. Two main categories were distinguished based on the use of Single- and Multi-materials. Their achieved mechanical properties were assessed by the comparison among MJ lattices and existing materials in Ashby plots. Thereby, the potential of MJ structures was described in terms of Young's modulus, yield strength and density values, which approached those regarding biomaterials and high-performing elastomers. Similarly, ranges of toughness close to the exhibited by foams were reported. These facts confirmed that biomedical and energy absorption applications were the most achieved design targets by MJ lattices (summary in Appendix~\ref{summary:AL}, Table~\ref{tab:SM}). Future works can start with the obtained outcomes from the exposed experimental and numerical results, and then try to cover existing gaps in the Ashby plots, mainly in areas nearby foams and biological materials. Shape programmable materials or vibration control devices are other promising targets that demand further research. The challenges and limitations of the MJ technique for lattice manufacturing have also been analyzed. The influence of UV intensity during the curing stage on the final mechanical performances and aging effects for the fabrication of long-term photo-polymeric lattice materials are areas that require more experimental work. Furthermore, high-cost estimations have mostly restricted the use of MJ in the AM market. Consequently, there is a lower number of research involved with respect to other technologies (\textit{e.g. FDM, EBM, SLS, SLA}). Despite the high printing cost, the MJ technique offers high resolution, reduced level of anisotropy in the printed components, and the capability of combining different materials to attain variable voxel-based mechanical properties.  Moreover, MJ is cost effective when it is printing parts in series or at full tray capacity. The use of materials with a controlled thermal response and different mechanical properties, that can be printed in a single round, is another remarkable MJ feature. These attributes pave the way for new research works to harness the potentiality of printing lattices with versatile photopolymers via MJ. At the same time, it can motivate a wider use of curable resins but not leaving aside the development of more sustainable alternatives, such as novel bio-resins.

\section{Funding}
SM, NMP, and DM are supported by the European Commission under the H2020 FET Open (‘‘Boheme’’) grant No. 863179 and by the Italian Ministry of Education, University and Research (MIUR) under the “Departments of Excellence” grant L. 232/2016. DM gratefully acknowledges financial support from ERC-ADG-2021-101052956-BEYOND.

\bibliography{references}

\begin{thebibliography}{100}
\expandafter\ifx\csname url\endcsname\relax
  \def\url#1{\texttt{#1}}\fi
\expandafter\ifx\csname urlprefix\endcsname\relax\def\urlprefix{URL }\fi
\expandafter\ifx\csname href\endcsname\relax
  \def\href#1#2{#2} \def\path#1{#1}\fi

\bibitem{Statista}
Statista,
  {\href{https://www.statista.com/statistics/756690/worldwide-most-used-3d-printing-technologies/}{Worldwide
  most used 3D printing technologies as of July 2018*}} (2022).

\bibitem{Stratasys2}
Stratasys, Empower medical innovation, Materials (2018).

\bibitem{3DsystemsPrototyping}
3D\hspace{2pt}Systems,
  {\href{http://it.infocenter.3dsystems.com/bestpractices/mjp-best-practices/projet-mjp-2500/general-multijet-printing/concept-modeling-rapid-prototyping}{Concept
  Modeling, Rapid Prototyping, Indirect Manufacturing}} (2020).

\bibitem{Stratasys1}
Stratasys,
  {\href{https://www.stratasys.com/en/3d-printers/printer-catalog/polyjet/?filter=PolyJet}{What
  is PolyJet Technology?}} (2022).

\bibitem{daminabo2020fused}
S.~C. Daminabo, S.~Goel, S.~A. Grammatikos, H.~Y. Nezhad, V.~K. Thakur, Fused
  deposition modeling-based additive manufacturing (3d printing): techniques
  for polymer material systems, Materials today chemistry 16 (2020) 100248.

\bibitem{gardner2020testing}
L.~Gardner, P.~Kyvelou, G.~Herbert, C.~Buchanan, Testing and initial
  verification of the world's first metal 3d printed bridge, Journal of
  constructional steel research 172 (2020) 106233.

\bibitem{vogel1995better}
S.~Vogel, Better bent than broken, Discover 16~(5) (1995) 62--67.

\bibitem{tofail2018additive}
S.~A. Tofail, E.~P. Koumoulos, A.~Bandyopadhyay, S.~Bose, L.~O’Donoghue,
  C.~Charitidis, Additive manufacturing: scientific and technological
  challenges, market uptake and opportunities, Materials today 21~(1) (2018)
  22--37.

\bibitem{rafiee2020multi}
M.~Rafiee, R.~D. Farahani, D.~Therriault, Multi-material 3d and 4d printing: a
  survey, Advanced Science 7~(12) (2020) 1902307.

\bibitem{3DSystemsintro}
3D\hspace{2pt}Systems,
  {\href{https://www.3dsystems.com/multi-jet-printing}{MultiJet Printing}}
  (2022).

\bibitem{Sculpteo}
Statista,
  {\href{https://www.statista.com/statistics/560304/worldwide-survey-3d-printing-top-technologies/}{Most
  used 3D printing technologies worldwide 2021}} (2022).

\bibitem{gulcan2021state}
O.~G{\"u}lcan, K.~G{\"u}nayd{\i}n, A.~Tamer, The state of the art of material
  jetting—a critical review, Polymers 13~(16) (2021) 2829.

\bibitem{3DsystemsBrochure}
3D\hspace{2pt}Systems,
  {\href{https://www.3dsystems.com/multi-jet-printing}{MultiJet Plastic
  Printers Brochure}} (2020).

\bibitem{quan2020photo}
H.~Quan, T.~Zhang, H.~Xu, S.~Luo, J.~Nie, X.~Zhu, Photo-curing 3d printing
  technique and its challenges, Bioactive Materials 5~(1) (2020) 110--115.

\bibitem{zabti2012effects}
M.~M. Zabti, Effects of light absorber on micro stereolithography parts, Ph.D.
  thesis, University of Birmingham (2012).

\bibitem{blanco2014nonisotropic}
D.~Blanco, P.~Fernandez, A.~Noriega, Nonisotropic experimental characterization
  of the relaxation modulus for polyjet manufactured parts, Journal of
  Materials Research 29~(17) (2014) 1876--1882.

\bibitem{gay2015analysis}
P.~Gay, D.~Blanco, F.~Pelayo, A.~Noriega, P.~Fern{\'a}ndez, Analysis of factors
  influencing the mechanical properties of flat polyjet manufactured parts,
  Procedia Engineering 132 (2015) 70--77.

\bibitem{tan2020recent}
L.~J. Tan, W.~Zhu, K.~Zhou, Recent progress on polymer materials for additive
  manufacturing, Advanced Functional Materials 30~(43) (2020) 2003062.

\bibitem{wagner2019cure}
A.~Wagner, I.~Gouzman, N.~Atar, E.~Grossman, M.~Pokrass, A.~Fuchsbauer,
  L.~Schranzhofer, C.~Paulik, Cure kinetics of bismaleimides as basis for
  polyimide-like inks for polyjet™-3d-printing, Journal of Applied Polymer
  Science 136~(12) (2019) 47244.

\bibitem{StratasysSUP}
Stratasys,
  {\href{https://support.stratasys.com/en/materials/polyjet/polyjet-support}{SUP706
  / SUP706B Support Material}} (2020).

\bibitem{Nahumsupport}
A.~Nahum,
  {\href{https://grabcad.com/tutorials/polyjet-support-materials-sup705-sup706b}{PolyJet
  Support Materials: SUP705 \& SUP706B}} (2020).

\bibitem{3Dsystemsclean}
3D\hspace{2pt}Systems,
  {\href{http://infocenter.3dsystems.com/projetmjp2500/post-processing-guide/visijet\%C2\%AE-materials/post-processing-part}{Post-Processing
  a part}} (2020).

\bibitem{ashby2006properties}
M.~F. Ashby, The properties of foams and lattices, Philosophical Transactions
  of the Royal Society A: Mathematical, Physical and Engineering Sciences
  364~(1838) (2006) 15--30.

\bibitem{Stratasys3}
Stratasys,
  {\href{https://www.stratasys.com/en/materials/search?technologies=731e07a1a51b42419acd1cb75142dfe6&sortIndex=0}{Materials
  Data Sheets}} (2022).

\bibitem{3DSystems}
3D\hspace{2pt}Systems, Multijet plastic printers visijet ® m3 advanced
  plastics (2017).

\bibitem{Varotsis}
A.~B. Varotsis,
  {\href{https://www.3dhubs.com/knowledge-base/introduction-material-jetting-3d-printing/}{Introduction
  to material jetting 3D printing}} (2020).

\bibitem{edgar2015additive}
J.~Edgar, S.~Tint, Additive manufacturing technologies: 3d printing, rapid
  prototyping, and direct digital manufacturing, Johnson Matthey Technology
  Review 59~(3) (2015) 193--198.

\bibitem{shapira2017next}
P.~Shapira, J.~Youtie, The next production revolution and institutions for
  technology diffusion, The Next Production Revolution: Implications for
  Governments and Business (2017).

\bibitem{PIERAU2022101517}
L.~Pierau, C.~Elian, J.~Akimoto, Y.~Ito, S.~Caillol, D.-L. Versace, Bio-sourced
  monomers and cationic photopolymerization–the green combination towards
  eco-friendly and non-toxic materials, Progress in Polymer Science 127 (2022)
  101517.

\bibitem{voet2021sustainable}
V.~S. Voet, J.~Guit, K.~Loos, Sustainable photopolymers in 3d printing: A
  review on biobased, biodegradable, and recyclable alternatives,
  Macromolecular Rapid Communications 42~(3) (2021) 2000475.

\bibitem{ChiulanPhotopolymerization}
I.~Chiulan, E.~B. Heggset, S.~I. Voicu, G.~Chinga-Carrasco, Photopolymerization
  of bio-based polymers in a biomedical engineering perspective,
  Biomacromolecules 22~(5) (2021) 1795--1814.

\bibitem{ashby1997cellular}
M.~F. Ashby, L.~J. Gibson, Cellular solids: structure and properties, Press
  Syndicate of the University of Cambridge, Cambridge, UK (1997) 175--231.

\bibitem{tao2016design}
W.~Tao, M.~C. Leu, Design of lattice structure for additive manufacturing, in:
  2016 International Symposium on Flexible Automation (ISFA), IEEE, 2016, pp.
  325--332.

\bibitem{tamburrino2018design}
F.~Tamburrino, S.~Graziosi, M.~Bordegoni, The design process of additively
  manufactured mesoscale lattice structures: a review, Journal of Computing and
  Information Science in Engineering 18~(4) (2018).

\bibitem{saxena2016three}
K.~K. Saxena, R.~Das, E.~P. Calius, Three decades of auxetics research-
  materials with negative poisson's ratio: a review, Advanced Engineering
  Materials 18~(11) (2016) 1847--1870.

\bibitem{lakes2017negative}
R.~S. Lakes, Negative-poisson's-ratio materials: auxetic solids, Annual review
  of materials research 47 (2017) 63--81.

\bibitem{kolken2017auxetic}
H.~M. Kolken, A.~Zadpoor, Auxetic mechanical metamaterials, RSC advances 7~(9)
  (2017) 5111--5129.

\bibitem{wang2016lightweight}
Q.~Wang, J.~A. Jackson, Q.~Ge, J.~B. Hopkins, C.~M. Spadaccini, N.~X. Fang,
  Lightweight mechanical metamaterials with tunable negative thermal expansion,
  Physical review letters 117~(17) (2016) 175901.

\bibitem{wu2019mechanical}
W.~Wu, W.~Hu, G.~Qian, H.~Liao, X.~Xu, F.~Berto, Mechanical design and
  multifunctional applications of chiral mechanical metamaterials: A review,
  Materials \& Design 180 (2019) 107950.

\bibitem{lim20192d}
T.-C. Lim, 2d metamaterial with in-plane positive and negative thermal
  expansion and thermal shearing based on interconnected alternating
  bimaterials, Materials Research Express 6~(11) (2019) 115804.

\bibitem{smith2004metamaterials}
D.~R. Smith, J.~B. Pendry, M.~C. Wiltshire, Metamaterials and negative
  refractive index, Science 305~(5685) (2004) 788--792.

\bibitem{PADILLA200628}
D.~R.~S. Willie J.~Padilla, Dimitri N.~Basov, Negative refractive index
  metamaterials, Materials Today 9~(7) (2006) 28--35.

\bibitem{Shalaev2007}
V.~M. Shalaev, Optical negative-index metamaterials, Nature Photonics 1 (2007)
  47--48.

\bibitem{PhysRevLett.102.023901}
S.~Zhang, Y.-S. Park, J.~Li, X.~Lu, W.~Zhang, X.~Zhang, Negative refractive
  index in chiral metamaterials, Phys. Rev. Lett. 102 (2009) 023901.

\bibitem{valdevit2016fabrication}
L.~Valdevit, J.~Bauer, Fabrication of 3d micro-architected/nano-architected
  materials, in: Three-dimensional microfabrication using two-photon
  polymerization, Elsevier, 2016, pp. 345--373.

\bibitem{jia2020engineering}
Z.~Jia, F.~Liu, X.~Jiang, L.~Wang, Engineering lattice metamaterials for
  extreme property, programmability, and multifunctionality, Journal of Applied
  Physics 127~(15) (2020) 150901.

\bibitem{valdevit2018architected}
L.~Valdevit, K.~Bertoldi, J.~Guest, C.~Spadaccini, Architected materials:
  synthesis, characterization, modeling, and optimal design (2018).

\bibitem{woesz2004cellular}
A.~Woesz, J.~Stampfl, P.~Fratzl, Cellular solids beyond the apparent
  density--an experimental assessment of mechanical properties, Advanced
  Engineering Materials 6~(3) (2004) 134--138.

\bibitem{fleck2010micro}
N.~A. Fleck, V.~S. Deshpande, M.~F. Ashby, Micro-architectured materials: past,
  present and future, Proceedings of the Royal Society A: Mathematical,
  Physical and Engineering Sciences 466~(2121) (2010) 2495--2516.

\bibitem{ai2017metamaterials}
L.~Ai, X.-L. Gao, Metamaterials with negative poisson’s ratio and
  non-positive thermal expansion, Composite Structures 162 (2017) 70--84.

\bibitem{ai2018three}
L.~Ai, X.-L. Gao, Three-dimensional metamaterials with a negative poisson's
  ratio and a non-positive coefficient of thermal expansion, International
  Journal of Mechanical Sciences 135 (2018) 101--113.

\bibitem{fleck2007damage}
N.~A. Fleck, X.~Qiu, The damage tolerance of elastic--brittle, two-dimensional
  isotropic lattices, Journal of the Mechanics and Physics of Solids 55~(3)
  (2007) 562--588.

\bibitem{fan2008yield}
H.~Fan, D.~Fang, F.~Jing, Yield surfaces and micro-failure mechanism of block
  lattice truss materials, Materials \& Design 29~(10) (2008) 2038--2042.

\bibitem{deshpande2001foam}
V.~Deshpande, M.~Ashby, N.~Fleck, Foam topology: bending versus stretching
  dominated architectures, Acta materialia 49~(6) (2001) 1035--1040.

\bibitem{kudo2019compressive}
A.~Kudo, D.~Misseroni, Y.~Wei, F.~Bosi, Compressive response of non-slender
  octet carbon microlattices, Frontiers in Materials 6 (2019) 169.

\bibitem{ambekar2021topologically}
R.~S. Ambekar, B.~Kushwaha, P.~Sharma, F.~Bosia, M.~Fraldi, N.~M. Pugno, C.~S.
  Tiwary, Topologically engineered 3d printed architectures with superior
  mechanical strength, Materials Today 48 (2021) 72--94.

\bibitem{barclift2012examining}
M.~W. Barclift, C.~B. Williams, Examining variability in the mechanical
  properties of parts manufactured via polyjet direct 3d printing, in: 2012
  International Solid Freeform Fabrication Symposium, University of Texas at
  Austin, 2012.

\bibitem{dizon2018mechanical}
J.~R.~C. Dizon, A.~H. Espera~Jr, Q.~Chen, R.~C. Advincula, Mechanical
  characterization of 3d-printed polymers, Additive Manufacturing 20 (2018)
  44--67.

\bibitem{kazmer2017three}
D.~Kazmer, Three-dimensional printing of plastics, in: Applied plastics
  engineering handbook, Elsevier, 2017, pp. 617--634.

\bibitem{maroti2018printing}
P.~Maroti, P.~Varga, H.~Abraham, G.~Falk, T.~Zsebe, Z.~Meiszterics, S.~Mano,
  Z.~Csernatony, S.~Rendeki, M.~Nyitrai, Printing orientation defines
  anisotropic mechanical properties in additive manufacturing of upper limb
  prosthetics, Materials Research Express 6~(3) (2018) 035403.

\bibitem{bochnia2016anisotrophy}
J.~Bochnia, S.~Blasiak, Anisotrophy of mechanical properties of a material
  which is shaped incrementally using polyjet technology, Engineering Mechanics
  2016 (2016) 74--77.

\bibitem{ulu2015enhancing}
E.~Ulu, E.~Korkmaz, K.~Yay, O.~Burak~Ozdoganlar, L.~Burak~Kara, Enhancing the
  structural performance of additively manufactured objects through build
  orientation optimization, Journal of Mechanical Design 137~(11) (2015).

\bibitem{konigshofer2021mechanical}
M.~K{\"o}nigshofer, M.~Stoiber, E.~Unger, C.~Grasl, F.~Moscato, Mechanical and
  dimensional investigation of additive manufactured multi-material parts,
  Frontiers in Physics 9 (2021) 128.

\bibitem{das2018effect}
S.~C. Das, R.~Ranganathan, N.~Murugan, Effect of build orientation on the
  strength and cost of polyjet 3d printed parts, Rapid Prototyping Journal
  (2018).

\bibitem{mueller2015tensile}
J.~Mueller, S.~E. Kim, K.~Shea, C.~Daraio, Tensile properties of inkjet 3d
  printed parts: critical process parameters and their efficient analysis, in:
  International Design Engineering Technical Conferences and Computers and
  Information in Engineering Conference, Vol. 57045, American Society of
  Mechanical Engineers, 2015, p. V01AT02A040.

\bibitem{cazon2014polyjet}
A.~Caz{\'o}n, P.~Morer, L.~Matey, Polyjet technology for product prototyping:
  Tensile strength and surface roughness properties, Proceedings of the
  Institution of Mechanical Engineers, Part B: Journal of Engineering
  Manufacture 228~(12) (2014) 1664--1675.

\bibitem{egan2019mechanics}
P.~F. Egan, I.~Bauer, K.~Shea, S.~J. Ferguson, Mechanics of three-dimensional
  printed lattices for biomedical devices, Journal of Mechanical Design 141~(3)
  (2019).

\bibitem{mueller2018buckling}
J.~Mueller, K.~Shea, Buckling, build orientation, and scaling effects in 3d
  printed lattices, Materials Today Communications 17 (2018) 69--75.

\bibitem{dalaq2016mechanical}
A.~S. Dalaq, D.~W. Abueidda, R.~K.~A. Al-Rub, Mechanical properties of 3d
  printed interpenetrating phase composites with novel architectured 3d
  solid-sheet reinforcements, Composites Part A: Applied Science and
  Manufacturing 84 (2016) 266--280.

\bibitem{vu2018characterizing}
I.~Q. Vu, L.~B. Bass, C.~B. Williams, D.~A. Dillard, Characterizing the effect
  of print orientation on interface integrity of multi-material jetting
  additive manufacturing, Additive Manufacturing 22 (2018) 447--461.

\bibitem{salcedo2018simulation}
E.~Salcedo, D.~Baek, A.~Berndt, J.~E. Ryu, Simulation and validation of three
  dimension functionally graded materials by material jetting, Additive
  Manufacturing 22 (2018) 351--359.

\bibitem{berselli2011hyperelastic}
G.~Berselli, R.~Vertechy, M.~Pellicciari, G.~Vassura, Hyperelastic modeling of
  rubber-like photopolymers for additive manufacturing processes, Rapid
  prototyping technology—Principles and functional requirements (2011)
  135--152.

\bibitem{tee2020polyjet}
Y.~L. Tee, C.~Peng, P.~Pille, M.~Leary, P.~Tran, Polyjet 3d printing of
  composite materials: experimental and modelling approach, Jom 72~(3) (2020)
  1105--1117.

\bibitem{slesarenko2018towards}
V.~Slesarenko, S.~Rudykh, Towards mechanical characterization of soft digital
  materials for multimaterial 3d-printing, International Journal of Engineering
  Science 123 (2018) 62--72.

\bibitem{zhang2016transversely}
P.~Zhang, A.~C. To, Transversely isotropic hyperelastic-viscoplastic model for
  glassy polymers with application to additive manufactured photopolymers,
  International Journal of Plasticity 80 (2016) 56--74.

\bibitem{liu2020maximizing}
W.~Liu, H.~Song, C.~Huang, Maximizing mechanical properties and minimizing
  support material of polyjet fabricated 3d lattice structures, Additive
  Manufacturing 35 (2020) 101257.

\bibitem{chen20183d}
Y.~Chen, T.~Li, Z.~Jia, F.~Scarpa, C.-W. Yao, L.~Wang, 3d printed hierarchical
  honeycombs with shape integrity under large compressive deformations,
  Materials \& Design 137 (2018) 226--234.

\bibitem{li2021additive}
T.~Li, F.~Jarrar, R.~A. Al-Rub, W.~Cantwell, Additive manufactured semi-plate
  lattice materials with high stiffness, strength and toughness, International
  Journal of Solids and Structures 230 (2021) 111153.

\bibitem{mohammadi2020hybrid}
K.~Mohammadi, M.~R. Movahhedy, I.~Shishkovsky, R.~Hedayati, Hybrid anisotropic
  pentamode mechanical metamaterial produced by additive manufacturing
  technique, Applied Physics Letters 117~(6) (2020) 061901.

\bibitem{afshar2018compressive}
M.~Afshar, A.~P. Anaraki, H.~Montazerian, Compressive characteristics of
  radially graded porosity scaffolds architectured with minimal surfaces,
  Materials Science and Engineering: C 92 (2018) 254--267.

\bibitem{zhang2021harnessing}
Z.~Zhang, F.~Scarpa, B.~A. Bednarcyk, Y.~Chen, Harnessing fractal cuts to
  design robust lattice metamaterials for energy dissipation, Additive
  Manufacturing 46 (2021) 102126.

\bibitem{white2021interpenetrating}
B.~C. White, A.~Garland, R.~Alberdi, B.~L. Boyce, Interpenetrating lattices
  with enhanced mechanical functionality, Additive Manufacturing 38 (2021)
  101741.

\bibitem{kumar2019tunable}
S.~Kumar, J.~Ubaid, R.~Abishera, A.~Schiffer, V.~Deshpande, Tunable energy
  absorption characteristics of architected honeycombs enabled via additive
  manufacturing, ACS applied materials \& interfaces 11~(45) (2019)
  42549--42560.

\bibitem{lancea2020compressive}
C.~Lancea, I.~Campbell, L.-A. Chicos, S.-M. Zaharia, Compressive behaviour of
  lattice structures manufactured by polyjet technologies, Polymers 12~(12)
  (2020) 2767.

\bibitem{sathishkumar2020mechanical}
N.~Sathishkumar, N.~Vivekanandan, L.~Balamurugan, N.~Arunkumar, I.~Ahamed,
  Mechanical properties of triply periodic minimal surface based lattices made
  by polyjet printing, Materials Today: Proceedings 22 (2020) 2934--2940.

\bibitem{kadkhodapour2014investigating}
J.~Kadkhodapour, H.~Montazerian, S.~Raeisi, Investigating internal architecture
  effect in plastic deformation and failure for tpms-based scaffolds using
  simulation methods and experimental procedure, Materials Science and
  Engineering: C 43 (2014) 587--597.

\bibitem{abate2020design}
K.~M. Abate, A.~Nazir, Y.-P. Yeh, J.-E. Chen, J.-Y. Jeng, Design, optimization,
  and validation of mechanical properties of different cellular structures for
  biomedical application, The International Journal of Advanced Manufacturing
  Technology 106~(3) (2020) 1253--1265.

\bibitem{alaboodi2018experimental}
A.~S. Alaboodi, S.~Sivasankaran, Experimental design and investigation on the
  mechanical behavior of novel 3d printed biocompatibility polycarbonate
  scaffolds for medical applications, Journal of Manufacturing Processes 35
  (2018) 479--491.

\bibitem{fernandes2021mechanically}
M.~C. Fernandes, J.~Aizenberg, J.~C. Weaver, K.~Bertoldi, Mechanically robust
  lattices inspired by deep-sea glass sponges, Nature Materials 20~(2) (2021)
  237--241.

\bibitem{janbaz2019ultra}
S.~Janbaz, F.~Bobbert, M.~Mirzaali, A.~Zadpoor, Ultra-programmable
  buckling-driven soft cellular mechanisms, Materials Horizons 6~(6) (2019)
  1138--1147.

\bibitem{weeger2019digital}
O.~Weeger, N.~Boddeti, S.-K. Yeung, S.~Kaijima, M.~L. Dunn, Digital design and
  nonlinear simulation for additive manufacturing of soft lattice structures,
  Additive Manufacturing 25 (2019) 39--49.

\bibitem{alberdi2020multi}
R.~Alberdi, R.~Dingreville, J.~Robbins, T.~Walsh, B.~C. White, B.~Jared, B.~L.
  Boyce, Multi-morphology lattices lead to improved plastic energy absorption,
  Materials \& Design 194 (2020) 108883.

\bibitem{mousanezhad2015hierarchical}
D.~Mousanezhad, S.~Babaee, H.~Ebrahimi, R.~Ghosh, A.~S. Hamouda, K.~Bertoldi,
  A.~Vaziri, Hierarchical honeycomb auxetic metamaterials, Scientific reports
  5~(1) (2015) 1--8.

\bibitem{guseinov2020programming}
R.~Guseinov, C.~McMahan, J.~P{\'e}rez, C.~Daraio, B.~Bickel, Programming
  temporal morphing of self-actuated shells, Nature communications 11~(1)
  (2020) 1--7.

\bibitem{johnson20183d}
L.~K. Johnson, C.~Richburg, M.~Lew, W.~R. Ledoux, P.~M. Aubin, E.~Rombokas, 3d
  printed lattice microstructures to mimic soft biological materials,
  Bioinspiration \& Biomimetics 14~(1) (2018) 016001.

\bibitem{jamshidian2020multiscale}
M.~Jamshidian, N.~Boddeti, D.~W. Rosen, O.~Weeger, Multiscale modelling of soft
  lattice metamaterials: Micromechanical nonlinear buckling analysis,
  experimental verification, and macroscale constitutive behaviour,
  International Journal of Mechanical Sciences 188 (2020) 105956.

\bibitem{Bertoldiannurev2017}
K.~Bertoldi, Harnessing instabilities to design tunable architected cellular
  materials, Annual Review of Materials Research 47~(1) (2017) 51--61.

\bibitem{papadopoulou2017auxetic}
A.~Papadopoulou, J.~Laucks, S.~Tibbits, Auxetic materials in design and
  architecture, Nature Reviews Materials 2~(12) (2017) 1--3.

\bibitem{ashjari2017auxetic}
M.~Ashjari, S.~Saadatm, S.~Hashemi, A.~Rasoulian, Auxetic materials materials
  with negative poisson’s ratio, Material Science \& Engineering
  International Journal 1~(2) (2017).

\bibitem{cui2018mechanical}
S.~Cui, B.~Gong, Q.~Ding, Y.~Sun, F.~Ren, X.~Liu, Q.~Yan, H.~Yang, X.~Wang,
  B.~Song, Mechanical metamaterials foams with tunable negative poisson’s
  ratio for enhanced energy absorption and damage resistance, Materials 11~(10)
  (2018) 1869.

\bibitem{jiang20203d}
H.~Jiang, Z.~Zhang, Y.~Chen, 3d printed tubular lattice metamaterials with
  engineered mechanical performance, Applied Physics Letters 117~(1) (2020)
  011906.

\bibitem{li2017harnessing}
T.~Li, X.~Hu, Y.~Chen, L.~Wang, Harnessing out-of-plane deformation to design
  3d architected lattice metamaterials with tunable poisson’s ratio,
  Scientific Reports 7~(1) (2017) 1--10.

\bibitem{chen2017lattice}
Y.~Chen, T.~Li, F.~Scarpa, L.~Wang, Lattice metamaterials with mechanically
  tunable poisson’s ratio for vibration control, Physical Review Applied
  7~(2) (2017) 024012.

\bibitem{wang2020design}
D.~Wang, H.~Xu, J.~Wang, C.~Jiang, X.~Zhu, Q.~Ge, G.~Gu, Design of 3d printed
  programmable horseshoe lattice structures based on a phase-evolution model,
  ACS Applied Materials \& Interfaces 12~(19) (2020) 22146--22156.

\bibitem{wagner2017large}
M.~Wagner, T.~Chen, K.~Shea, Large shape transforming 4d auxetic structures, 3D
  printing and Additive Manufacturing 4~(3) (2017) 133--142.

\bibitem{dong2021design}
L.~Dong, C.~Jiang, J.~Wang, D.~Wang, Design of shape reconfigurable, highly
  stretchable honeycomb lattice with tunable poisson’s ratio,
  Four-dimensional (4D) Printing (2021).

\bibitem{xie2018acoustic}
Y.~Xie, Y.~Fu, Z.~Jia, J.~Li, C.~Shen, Y.~Xu, H.~Chen, S.~A. Cummer, Acoustic
  imaging with metamaterial luneburg lenses, Scientific reports 8~(1) (2018)
  1--6.

\bibitem{yi20163d}
J.~Yi, S.~N. Burokur, G.-P. Piau, A.~de~Lustrac, 3d printed broadband
  transformation optics based all-dielectric microwave lenses, Journal of
  Optics 18~(4) (2016) 044010.

\bibitem{chen2017broadband}
Y.~Chen, F.~Qian, L.~Zuo, F.~Scarpa, L.~Wang, Broadband and multiband vibration
  mitigation in lattice metamaterials with sinusoidally-shaped ligaments,
  Extreme Mechanics Letters 17 (2017) 24--32.

\bibitem{lim2021phononic}
C.~Lim, et~al., Phononic metastructures with ultrawide low frequency
  three-dimensional bandgaps as broadband low frequency filter, Scientific
  reports 11~(1) (2021) 1--11.

\bibitem{peri2019axial}
V.~Peri, M.~Serra-Garcia, R.~Ilan, S.~D. Huber, Axial-field-induced chiral
  channels in an acoustic weyl system, Nature Physics 15~(4) (2019) 357--361.

\bibitem{vdovin2017implementation}
R.~Vdovin, T.~Tomilina, V.~Smelov, M.~Laktionova, Implementation of the
  additive polyjet technology to the development and fabricating the samples of
  the acoustic metamaterials, Procedia Engineering 176 (2017) 595--599.

\bibitem{mueller2017mechanical}
J.~Mueller, D.~Courty, M.~Spielhofer, R.~Spolenak, K.~Shea, Mechanical
  properties of interfaces in inkjet 3d printed single-and multi-material
  parts, 3D Printing and Additive Manufacturing 4~(4) (2017) 193--199.

\bibitem{gu2016biomimetic}
G.~X. Gu, M.~Takaffoli, A.~J. Hsieh, M.~J. Buehler, Biomimetic additive
  manufactured polymer composites for improved impact resistance, Extreme
  Mechanics Letters 9 (2016) 317--323.

\bibitem{zaheri2018revealing}
A.~Zaheri, J.~S. Fenner, B.~P. Russell, D.~Restrepo, M.~Daly, D.~Wang,
  C.~Hayashi, M.~A. Meyers, P.~D. Zavattieri, H.~D. Espinosa, Revealing the
  mechanics of helicoidal composites through additive manufacturing and beetle
  developmental stage analysis, Advanced Functional Materials 28~(33) (2018)
  1803073.

\bibitem{jia20193d}
Z.~Jia, L.~Wang, 3d printing of biomimetic composites with improved fracture
  toughness, Acta Materialia 173 (2019) 61--73.

\bibitem{yan2021recent}
X.~Yan, H.~Chen, S.~Lin, S.~Xiao, Y.~Yang, Recent advancements in biomimetic 3d
  printing materials with enhanced mechanical properties, Frontiers in
  Materials 8 (2021) 139.

\bibitem{li2018enhanced}
T.~Li, Y.~Chen, L.~Wang, Enhanced fracture toughness in architected
  interpenetrating phase composites by 3d printing, Composites Science and
  Technology 167 (2018) 251--259.

\bibitem{al2017mechanical}
O.~Al-Ketan, M.~A. Assad, R.~K.~A. Al-Rub, Mechanical properties of periodic
  interpenetrating phase composites with novel architected microstructures,
  Composite Structures 176 (2017) 9--19.

\bibitem{li2018exploiting}
T.~Li, Y.~Chen, X.~Hu, Y.~Li, L.~Wang, Exploiting negative poisson's ratio to
  design 3d-printed composites with enhanced mechanical properties, Materials
  \& design 142 (2018) 247--258.

\bibitem{albertini2021experimental}
F.~Albertini, J.~Dirrenberger, C.~Sollogoub, T.~Maconachie, M.~Leary,
  A.~Molotnikov, Experimental and computational analysis of the mechanical
  properties of composite auxetic lattice structures, Additive Manufacturing 47
  (2021) 102351.

\bibitem{janbaz2018multimaterial}
S.~Janbaz, M.~McGuinness, A.~A. Zadpoor, Multimaterial control of instability
  in soft mechanical metamaterials, Physical Review Applied 9~(6) (2018)
  064013.

\bibitem{mirzaali2018multi}
M.~Mirzaali, A.~Caracciolo, H.~Pahlavani, S.~Janbaz, L.~Vergani, A.~Zadpoor,
  Multi-material 3d printed mechanical metamaterials: Rational design of
  elastic properties through spatial distribution of hard and soft phases,
  Applied Physics Letters 113~(24) (2018) 241903.

\bibitem{jiang20183d}
Y.~Jiang, Y.~Li, 3d printed auxetic mechanical metamaterial with chiral cells
  and re-entrant cores, Scientific reports 8~(1) (2018) 1--11.

\bibitem{bossart2021oligomodal}
A.~Bossart, D.~M. Dykstra, J.~Van~der Laan, C.~Coulais, Oligomodal
  metamaterials with multifunctional mechanics, Proceedings of the National
  Academy of Sciences 118~(21) (2021).

\bibitem{hoberman1990reversibly}
C.~Hoberman, Reversibly expandable doubly-curved truss structure, uS Patent
  4,942,700 (Jul.~24 1990).

\bibitem{li2019phononic}
Y.~Li, S.~Cao, Y.~Shen, Y.~Meng, Phononic band-gaps of hoberman spherical
  metamaterials in low frequencies, Materials \& Design 181 (2019) 107935.

\bibitem{li2018hoberman}
Y.~Li, Y.~Chen, T.~Li, S.~Cao, L.~Wang, Hoberman-sphere-inspired lattice
  metamaterials with tunable negative thermal expansion, Composite Structures
  189 (2018) 586--597.

\bibitem{al2018nature}
O.~Al-Ketan, A.~Soliman, A.~M. AlQubaisi, R.~K. Abu Al-Rub, Nature-inspired
  lightweight cellular co-continuous composites with architected periodic
  gyroidal structures, Advanced Engineering Materials 20~(2) (2018) 1700549.

\bibitem{lumpe2021computational}
T.~S. Lumpe, K.~Shea, Computational design of 3d-printed active lattice
  structures for reversible shape morphing, Journal of Materials Research
  36~(18) (2021) 3642--3655.

\bibitem{hedayati20213d}
R.~Hedayati, A.~G{\"u}ven, S.~Van Der~Zwaag, 3d gradient auxetic soft
  mechanical metamaterials fabricated by additive manufacturing, Applied
  Physics Letters 118~(14) (2021) 141904.

\bibitem{mirzaali2020non}
M.~Mirzaali, H.~Pahlavani, E.~Yarali, A.~Zadpoor, Non-affinity in
  multi-material mechanical metamaterials, Scientific reports 10~(1) (2020)
  1--10.

\bibitem{mueller2018stepwise}
J.~Mueller, K.~Shea, Stepwise graded struts for maximizing energy absorption in
  lattices, Extreme Mechanics Letters 25 (2018) 7--15.

\bibitem{wang2015designable}
K.~Wang, Y.-H. Chang, Y.~Chen, C.~Zhang, B.~Wang, Designable dual-material
  auxetic metamaterials using three-dimensional printing, Materials \& Design
  67 (2015) 159--164.

\bibitem{saxena20173d}
K.~K. Saxena, R.~Das, E.~P. Calius, 3d printable multimaterial cellular
  auxetics with tunable stiffness, arXiv preprint arXiv:1707.04486 (2017).

\bibitem{wu2016isotropic}
L.~Wu, B.~Li, J.~Zhou, Isotropic negative thermal expansion metamaterials, ACS
  applied materials \& interfaces 8~(27) (2016) 17721--17727.

\bibitem{3Dsystemsbestpractice}
3D\hspace{2pt}Systems,
  {\href{http://it.infocenter.3dsystems.com/bestpractices/mjp-best-practices/projet-mjp-2500/general-multijet-printing/designing-shells-and-infill}{Best
  Practice-General MultiJet Printing-Designing with Shells and Infill}} (2022).

\bibitem{Stratasysguideline}
Stratasys,
  {\href{https://my.stratasys.com/SupportCenter/HTML5UserGuides/Design_DFAM_Guide_July_2020/Responsive\%20HTML5/DOC-01103_x_Design-PJ-AM-Guide-HTML/DfAM_Guide-Chapter/DfAM_Guide-Chapter.htm}{Design
  for Additive Manufacturing with PolyJet}} (2022).

\bibitem{baumers2015modeling}
M.~Baumers, R.~Wildman, C.~Tuck, P.~Dickens, R.~Hague, Modeling build time,
  process energy consumption and cost of material jetting-based additive
  manufacturing, in: NIP \& Digital Fabrication Conference, Vol. 2015, Society
  for Imaging Science and Technology, 2015, pp. 311--316.

\bibitem{fleckdeshpandewebinarEML}
N.~Fleck, V.~Deshpande, New horizons in microarchitectured materials, Extreme
  Mechanical Letters webinars (2020).

\bibitem{khaderi2017indentation}
S.~N. Khaderi, M.~Scherer, C.~Hall, U.~Steiner, U.~Ramamurty, N.~Fleck,
  V.~Deshpande, The indentation response of nickel nano double gyroid lattices,
  Extreme Mechanics Letters 10 (2017) 15--23.

\bibitem{ashbymaterials}
M.~F. Ashby, Materials selection in mechanical design, in: Materials selection
  in mechanical design, Vol.~4, Elsevier, 2011.

\bibitem{Granta_Design}
Granta\hspace{2pt}Design, Ces edupack 2019.

\bibitem{eckner2020investigations}
R.~Eckner, C.~Baumgart, L.~Kr{\"u}ger, Investigations on the influence of
  strain rate, temperature and reinforcement on strength and deformation
  behavior of crmnni-steels, in: Austenitic TRIP/TWIP Steels and Steel-Zirconia
  Composites, Springer, 2020, pp. 379--412.

\bibitem{haghpanah2016multistable}
B.~Haghpanah, L.~Salari-Sharif, P.~Pourrajab, J.~Hopkins, L.~Valdevit,
  Multistable shape-reconfigurable architected materials, Advanced Materials
  28~(36) (2016) 7915--7920.

\bibitem{mallick2012failure}
P.~Mallick, Failure of polymer matrix composites (pmcs) in automotive and
  transportation applications, in: Failure Mechanisms in Polymer Matrix
  Composites, Elsevier, 2012, pp. 368--392.

\bibitem{DassaultSystemes}
Dassault\hspace{2pt}Systemes,
  {\href{https://make.3dexperience.3ds.com/processes/introduction-to-additive-processes}{Introduction
  to 3D printing - additive processes}} (2020).

\bibitem{Stratasysj850}
Stratasys,
  {\href{https://www.stratasys.com/contentassets/13997c2749194823be3f37f121311357/br_pj_me_j850-digital-anatomy_a4_0122a-2.pdf?v=495b7c}{J850™
  Digital Anatomy Printer Solutions}} (2022).

\bibitem{meisel2014design}
N.~A. Meisel, C.~B. Williams, Design for additive manufacturing: an
  investigation of key manufacturing considerations in multi-material polyjet
  3d printing, in: 2014 International Solid Freeform Fabrication Symposium,
  University of Texas at Austin, 2014.

\bibitem{StratasysSLAvsPJ}
Stratasys,
  {\href{https://www.stratasysdirect.com/manufacturing-services/3d-printing/differences-between-stereolithography-polyjet}{Top
  4 Differences Between Stereolithography and PolyJet}} (2022).

\bibitem{bennett2017measuring}
J.~Bennett, Measuring uv curing parameters of commercial photopolymers used in
  additive manufacturing, Additive manufacturing 18 (2017) 203--212.

\bibitem{jacobs1996stereolithography}
P.~Jacobs, Stereolithography and other rp\&m technologies. society of
  manufacturing engineers, New York (1996) 1--26.

\bibitem{pandey2014photopolymers}
R.~Pandey, Photopolymers in 3d printing applications (2014).

\bibitem{layani2018novel}
M.~Layani, X.~Wang, S.~Magdassi, Novel materials for 3d printing by
  photopolymerization, Advanced Materials 30~(41) (2018) 1706344.

\bibitem{bagheri2019photopolymerization}
A.~Bagheri, J.~Jin, Photopolymerization in 3d printing, ACS Applied Polymer
  Materials 1~(4) (2019) 593--611.

\bibitem{StratasysHighT}
Stratasys,
  {\href{https://www.stratasys.com/en/materials/materials-catalog/polyjet-materials/high-temperature/}{High
  Temperature}} (2022).

\bibitem{3DSystemsHighT}
3D\hspace{2pt}Systems,
  {\href{https://www.3dsystems.com/materials/visijet-m2s-ht250}{VisiJet
  M2S-HT250 (MJP)}} (2022).

\bibitem{3DSystemsMJP5600}
3D\hspace{2pt}Systems,
  {\href{https://pdf.directindustry.com/pdf/3d-systems/projet-mjp-5600/19418-724109-_2.html}{VisiJet
  Base Materials for the ProJet MJP 5600}} (2017).

\bibitem{meisel2018impact}
N.~A. Meisel, D.~A. Dillard, C.~B. Williams, Impact of material concentration
  and distribution on composite parts manufactured via multi-material jetting,
  Rapid Prototyping Journal (2018).

\bibitem{khoshkhoo2018effect}
A.~Khoshkhoo, A.~L. Carrano, D.~M. Blersch, Effect of build orientation and
  part thickness on dimensional distortion in material jetting processes, Rapid
  Prototyping Journal (2018).

\bibitem{tee20203d}
Y.~L. Tee, P.~Tran, M.~Leary, P.~Pille, M.~Brandt, 3d printing of polymer
  composites with material jetting: Mechanical and fractographic analysis,
  Additive Manufacturing 36 (2020) 101558.

\bibitem{sugavaneswaran2015analytical}
M.~Sugavaneswaran, G.~Arumaikkannu, Analytical and experimental investigation
  on elastic modulus of reinforced additive manufactured structure, Materials
  \& Design (1980-2015) 66 (2015) 29--36.

\bibitem{bass2016exploring}
L.~Bass, N.~A. Meisel, C.~B. Williams, Exploring variability of orientation and
  aging effects in material properties of multi-material jetting parts, Rapid
  Prototyping Journal (2016).

\bibitem{safai2019review}
L.~Safai, J.~S. Cuellar, G.~Smit, A.~A. Zadpoor, A review of the fatigue
  behavior of 3d printed polymers, Additive manufacturing 28 (2019) 87--97.

\bibitem{yap2020review}
Y.~L. Yap, S.~L. Sing, W.~Y. Yeong, A review of 3d printing processes and
  materials for soft robotics, Rapid Prototyping Journal (2020).

\bibitem{moore2012fatigue}
J.~P. Moore, C.~B. Williams, Fatigue characterization of 3d printed elastomer
  material, in: 2012 International Solid Freeform Fabrication Symposium,
  University of Texas at Austin, 2012.

\bibitem{pugalendhi2020effect}
A.~Pugalendhi, R.~Ranganathan, M.~Chandrasekaran, Effect of process parameters
  on mechanical properties of veroblue material and their optimal selection in
  polyjet technology, The International Journal of Advanced Manufacturing
  Technology 108~(4) (2020) 1049--1059.

\bibitem{moore2015fatigue}
J.~P. Moore, C.~B. Williams, Fatigue properties of parts printed by polyjet
  material jetting, Rapid Prototyping Journal (2015).

\bibitem{suresh2018fatigue}
J.~Suresh, G.~Saravana~Kumar, P.~Ramu, J.~Rengaswamy, Fatigue life
  characterization of additively manufactured acrylic like poly-jet printed
  parts, in: Advances in Structural Integrity, Springer, 2018, pp. 623--632.

\bibitem{zheng2014ultralight}
X.~Zheng, H.~Lee, T.~H. Weisgraber, M.~Shusteff, J.~DeOtte, E.~B. Duoss, J.~D.
  Kuntz, M.~M. Biener, Q.~Ge, J.~A. Jackson, et~al., Ultralight, ultrastiff
  mechanical metamaterials, Science 344~(6190) (2014) 1373--1377.

\bibitem{surjadi2019mechanical}
J.~U. Surjadi, L.~Gao, H.~Du, X.~Li, X.~Xiong, N.~X. Fang, Y.~Lu, Mechanical
  metamaterials and their engineering applications, Advanced Engineering
  Materials 21~(3) (2019) 1800864.

\bibitem{nguyen2021design}
C.~H.~P. Nguyen, Y.~Kim, Y.~Choi, Design for additive manufacturing of
  functionally graded lattice structures: A design method with process induced
  anisotropy consideration, International Journal of Precision Engineering and
  Manufacturing-Green Technology 8~(1) (2021) 29--45.

\bibitem{boniotti2019analysis}
L.~Boniotti, S.~Foletti, S.~Beretta, L.~Patriarca, Analysis of strain and
  stress concentrations in micro-lattice structures manufactured by slm, Rapid
  Prototyping Journal (2019).

\bibitem{zhang2020design}
X.~Zhang, Y.~Wang, B.~Ding, X.~Li, Design, fabrication, and mechanics of 3d
  micro-/nanolattices, Small 16~(15) (2020) 1902842.

\bibitem{li2017cost}
Y.~Li, B.~S. Linke, H.~Voet, B.~Falk, R.~Schmitt, M.~Lam, Cost, sustainability
  and surface roughness quality--a comprehensive analysis of products made with
  personal 3d printers, CIRP Journal of Manufacturing Science and Technology 16
  (2017) 1--11.

\bibitem{Prusa}
Prusa,
  {\href{https://www.prusa3d.com/product/original-prusa-i3-mk3s-3d-printer-3/}{Original
  Prusa i3 MK3S + 3D printer}} (2022).

\bibitem{PrusaPCblend}
Prusa,
  {\href{https://www.prusa3d.com/product/prusament-pc-blend-jet-black-970g/}{Prusament
  PC Blend Jet Black 970g}} (2022).

\bibitem{PrusaSupport}
Prusa,
  {\href{https://www.prusa3d.com/product/verbatim-bvoh-soluble-support-500g/}{Verbatim
  BVOH Soluble Support 500g}} (2022).

\end{thebibliography}

\begin{appendices}
\section{MJ additional remarks}\label{app_MJ}
\subsection{MJ technology}\label{app_MJprocess}
Material jetting (MJ)  involves the following main groups: \textit{Drop On Demand (DOD), Nano Particle Jetting (NPJ),  PolyJet (PJ)~\citep{DassaultSystemes}, MultiJet Printing (MJP)}. In this review, we focus on these two last techniques that involve photopolymers, and both are the most widely utilized to manufacture automotive, aerospace, and medical components. Polyjet$^\text{\textregistered}$~\citep{Stratasys1} and MultiJet$^\text{\textregistered}$~\citep{3DSystemsintro} present advantages versus other 3D printing technologies in terms of precision, uniformity, and very well detailed surfaces. Current MJ 3D printers fabricate parts in a high resolution depending on their available printing modes. For example, minimum layers height of 13 and 14 microns can be obtained by selecting Extreme High definition (XHD)~\citep{3DsystemsBrochure} and High Quality (HQ) modes~\citep{Stratasysj850}, respectively. In addition, it is possible to choose printing modes that work almost two times faster than the latter ones, such as High-Mix (layers height of 27 microns)~\citep{Stratasysj850} or High Definition (layers height of 32 microns)~\citep{3DsystemsBrochure}.  The schematic of the working principle behind the MJ process is depicted in Fig.\ref{fig:01}(B). During the MJ process, the photopolymer droplets are spread and deposited as small voxels by multiple nozzles onto a horizontal build tray, similar to inkjet printers. The deposed substrates are cured instantaneously by UV light. Depending on the model's size amount, finishing and accuracy, a support material grid is also printed to fill overhangs, gaps, and spaces  during the layers' build-up process, as it is schematized in Fig.~\ref{fig:01}(B). In the case of Polyjet, the soluble gel support material is printed surrounding and withstanding the floating parts by forming a grid structure. The strength of the supports is controlled by the grid's density and the inclusion of rigid base material particles in the support matrix. It can vary between lite, standard, and heavy levels, depending on the geometry and material to be held. Parts with angled faces almost perpendicular or inclined approximately above 72° with respect to the build tray, may be self-supported without the use of support material grid~\citep{meisel2014design}. The removal operations are usually done by water pressure or alkaline solutions for difficult-to-reach areas and fragile models. In general, the supports become completed melted by 2\% Na(OH) solutions after an extended time (usually less than 4 hours by using a cleaning station and in the case of lite support). Alkaline solutions demand special attention because they can have adverse effects on the mechanical properties of the final 3D printed part~\citep{Nahumsupport}. However, it is suggested a brief submersion of the parts (30-60 seconds) in 15\% glycerol-based solutions to wash away the solution remainders.  MultiJet printing follows the same photopolymerization process~\citep{tan2020recent}. Nevertheless, the support material consists of wax and it can be removed by heat. For this reason, there is available equipment such as convection ovens (set to 65°C) or steamer systems that are able to melt the wax. Further removal of fine wax particles can be done also by ultrasonic bath devices~\citep{3Dsystemsclean}. 
The use of support material influences the reduction of precision along printed parts dimensions. MJ techniques and SLA share common curing principles. The difference is that the latter forms the targeted shape inside a resin pool. As the build tray moves up (on the vertical axis), the desired patterns are constructed by hardening the photopolymer via UV laser coming from dynamic mirrors. As the layers are formed, the objects get suspended, and then an additional support structure made of the same material is also needed. The final parts are not fully hardened because of the excess resin, and they require further curing in a UV source (e.g. oven) to eliminate it. Then, the supports can be removed by manual operations, including sanding ~\citep{StratasysSLAvsPJ}. The effect of UV curing and post-processing affect MJ materials' mechanical properties significantly, similarly to SLA~\citep{dizon2018mechanical}. The main parameters that govern photopolymerization are the critical energy that initiates the process ($E_{cr}$)  and the penetration depth of curing light ($D_{p}$)~\citep{bennett2017measuring}. Considering the energy of light at the surface ($E_{l}$), the thickness ($h_{c}$) at which the photopolymer is cured is equal to $h_{c}= (D_{p}\cdot \ln (E_{l}\div E_{cr}))$~\citep{jacobs1996stereolithography}. Whether the parameters $E_{cr}$ and $D_{p}$ are known, it is possible to determine the required number of increments to build-up the layer in the vertical axis to obtain the minimal thickness $h_{c}$ for layers deposition and curing, and the optimal UV light features (e.g. scan speed and intensity).  Thus, high printing resolution can be guaranteed~\citep{bennett2017measuring}. Moreover, the wavelength of the light source (e.g. lamp) in MJ printers is not constrained. For this reason, it is also possible to select a curing approach based on hybrid photopolymerization (radical and cationic)~\citep{quan2020photo}. Both MJ and SLA share similar photopolymerization challenges, such as assuring a fast and efficient curing rate that avoids differences in bonding and hardening, along and between the layers. Photoinitiators are an important component of the resins, they absorb the UV light energy and it is transformed into chemical energy. Free cations or radicals are formed  and they break into more particles that will interact and create bonds with the constituent monomers and oligomers~\citep{pandey2014photopolymers}. Their properties are formulated based on the wavelength and the intensity of the UV light source. They can be optimized in order to obtain high printing resolution~\citep{layani2018novel}. Besides the use of low viscosity resins, photocurable materials, that can be cured under longer wavelengths of irradiated UV light, have been proposed. This fact can also contribute to less dangerous working environments for health. Thereby, a higher penetration depth between neighboring layers can also be obtained and thus, achieving stronger bonds during the build-up process~\citep{bagheri2019photopolymerization}.

\subsection{Manufacturing process effects on mechanical properties of MJ printed parts}\label{app_MJprocess:mech}
One of the most outstanding features of MJ, is the use of multiple materials with different physical and mechanical characteristics that can be printed at the same time, as shown in Table~\ref{Table:3}. For this reason, they present feasible options to obtain versatile printed parts with rigid or flexible, multi-color or translucent components, as well as heat resistant (e.g. with heat deflection temperature of 67°C~\citep{StratasysHighT}, and 250°C~\citep{3DSystemsHighT} at about 0.45~MPa), bio-compatible, casting wax and compliant mechanisms (e.g. polypropylene alike-photopolymers)~\citep{Stratasys3,3DSystemsintro}. It is also possible to create a sort of composites with hybrid characteristics which are the result of the addition of rigid photopolymers to soft rubbery resins, such as the so-called \textit{"Digital Materials" (DM), "Multi-Material Composites" (MMC), or "Functionally Graded Materials" (FGM)}.
\begin{table*}[t]
 \centering
 	\caption{Mechanical properties of common Polyjet~\citep{Stratasys3} and MultiJet printing base materials~\citep{3DSystems,3DSystemsMJP5600}. ($^*$ Rubber like materials)}
 {\renewcommand{\arraystretch}{1.5}
 \begin{tabular}{*1{>{\centering\arraybackslash}m{0.02\textwidth}}*1{>{\centering\arraybackslash}m{0.25\textwidth}}*6{>{\centering\arraybackslash}m{0.07\textwidth}}}
\otoprule
& Material & Tensile Strength (MPa) & Elongation at break (\%) & Young's Modulus (MPa) & Flexural Strength (MPa) & Shore Hardness (Scale) & Density (g/cm3)\\
 \hline
\parbox[t]{2mm}{\multirow{10}{*}{\rotatebox[origin=c]{90}{\textbf{Polyjet}}}} &$^*$Agilus30-Black   & 2.4-3.1    &220-270 &  - & -    &30-35A &   1.14\\
& Digital ABSplus   & 55-60    &25-40 & 2600-3000& 65-75    &85-87D &   1.17\\
& Draft Gray   & 50-65    &10-25 & 2000-3000& 75-110    &83-86D &   1.17\\
& Durus   & 20-30    &40-50 & 1000-1200& 30-40    &74-78D &   1.15\\
& Rigur   & 40-65    &20-35 & 1700-2100& 52-59    &80-84D &   1.20\\
& $^*$TangoBlackPlus   & 0.8-1.5    &170-220 & - & -    &26-28A &   1.12\\
& Vero WhitePlus/Black/Magenta   & 50-65    &10-25 & 2000-3000& 75-110    &83-86D &   1.17\\
& VeroBlue   & 50-60    &15-25 & 2000-3000& 60-70    &83-86D &   1.18\\
& VeroClear   & 50-65    &10-25 & 2000-3000& 75-110    &83-86D &   1.18\\
& VeroUltraClear   & 39-43    &20-35 & 1400-2100& 58-72    &80-85D &   1.18\\
 \hline
\parbox[t]{2mm}{\multirow{10}{*}{\rotatebox[origin=c]{90}{\textbf{MultiJet}}}} 
& VisiJet CR-WT/ CR-CL   & 37-47    &7-16 & -& 61-72    & 76-80D &   1.18\\
& VisiJet CR-BK   & 45-52    &7-11 & -& 63-76    & 78-83D &   1.18\\
&  $^*$VisiJet CE-NT/BK   & 0.2-0.4    &160-230 & -& -    & 27-33A &   1.12\\
& VisiJet M3-X White   & 49    &8.3 & 2168& 65    & - &   1.04\\
& VisiJet M3 Crystal   & 42.4    &6.83 & 1463& 49    & - &   1.02\\
& VisiJet M3 Black   & 35.2    &19.7 & 1594& 44.5    & - &   1.02\\
& VisiJet M3 Proplast   & 26.2    &8.97 & 1108& 26.6    & - &   1.02\\
& VisiJet M3 Navy   & 20.5    &8 & 735& 28.1    & - &   1.02\\
& VisiJet M3 Techplast   & 22.1    &6.1 & 866& 28.1    & - &   1.02\\
&  VisiJet M3 Procast   & 32    &12.3 & 1724& 45    & - &   1.02\\

 \hline
 \end{tabular}}
 \label{Table:3}
 \end{table*}
Therefore, they offer a broad spectrum of properties to be explored, such as intermediate levels of stiffness or flexibility  with respect to the glassy or soft former photopolymers, as well as different level of shore hardness scale (from A30 to A95 for flexible and D70 to D85 for glassy materials), and viscoelastic properties that vary according to the distribution of rigid versus soft particles~\citep{Stratasys3,3DSystems,3DSystemsMJP5600,meisel2018impact}. In the case of rigid lattices made of materials like VeroWhite, the selection of compressive dominated lattices over bending dominated ones is recommended for improved strength or energy absorption values~\citep{zhang2016transversely}. 
Material characterization experiments on Polyjet printed specimens have been conducted mostly under common standards such as ASTM D-638 for tensile tests, ASTM D790 for flexural tests, or ASTM D2240 for shore hardness. The layers' orientation of the specimens with respect to the build tray, contributes to the anisotropy of the mechanical properties. Besides build-orientation, parameters such as struts buckling and scaling effects of printed lattices affect their final mechanical performances~\citep{mueller2018buckling}. Distortion would be present along the longest dimension of the part, but increments in their thickness might prevent higher deformations~\citep{khoshkhoo2018effect}. Moreover, the effect of the printing direction in 3D printed composites may be negligible until  5\% of volume fractions of reinforced materials~\citep{tee20203d}.
The allocation of rigid particles affects the mechanical behavior of the specimens. Those perpendicular to the force allow a higher compressive resistance. While in the case of tensile strength, the higher values were obtained by the parts aligned parallel to the print head. Soft particles can produce the cracks initiation in the formed composite~\citep{tee20203d,tee2020polyjet,sugavaneswaran2015analytical,vu2018characterizing,bass2016exploring}. In addition, aging effects have been initially observed in glassy photopolymers samples, where the ultimate tensile strength incremented in time but elongation at break values diminished~\citep{bass2016exploring}.
It's still challenging to identify which process parameters mainly affect the fatigue behavior of MJ parts~\citep{safai2019review,yap2020review}. Surface flaws, voids and shear forces action tend to decrease the fatigue life of elastomeric and multi-material interfaces~\citep{moore2012fatigue}. However, the type of surface finishing might improve the fatigue life as well as the mechanical strength of the printed part. In particular, glossy finishing surfaces have presented fewer irregularities that induce crack propagation than their matte counterparts~\citep{pugalendhi2020effect,cazon2014polyjet,moore2015fatigue}. Moreover, layers' orientation can affect the crack propagation within the printed components. Cracks formed orthogonal to the layers' orientation propagate slower than those formed parallel~\citep{suresh2018fatigue}.
Blocked nozzles tend to increase the surface roughness. Therefore, an increment in stress concentrations is expected~\citep{mueller2015tensile}. 
Higher layer thickness represents a time-saving and enhanced bond between the build-up substrates. 
Despite a resolution reduction, high-speed printing and thickness of 30 microns allowed an enhanced bond between layers and less time consumption (saving 60.86\%) with respect to high quality-matte models (thickness of 16~microns)~\citep{pugalendhi2020effect}.

\section{3D printed architected lattices: common manufacturing processes highlights}\label{app_MJ_lat}
\subsection{3D printing techniques}\label{app_MJ_printing:technique}
The latest developments in 3D printing techniques have allowed the possibility to produce complex and enhanced topologies in a wide-scale range from nano and microstructures, besides the fabrication at the macro-scale (e.g. long parts). The existing AM techniques, classified according to
ASTM-ISO standards, work with different processes, types of
raw materials, and printing resolution [8], as it is summarized
in Table~\ref{Table:3}. Micro-stereolithography has been one of the pioneer technologies whose resolution reaches micrometers. For example, structures can be printed with a resolution close to 5 microns~\citep{zheng2014ultralight}.
Furthermore, the main techniques used to print metamaterials at the microscale, among photopolymerization categories, are: Self Propagating Photopolymer Waveguide (SPPW), Two-Photon Lithography (TPL), Continuous Liquid Interface Production (CLIP)~\citep{surjadi2019mechanical}. In the case of metals, SLS, SLM, EBM, and DED have been widely used~\citep{surjadi2019mechanical,nguyen2021design}.

\subsection{Manufacturing effects on lattices}\label{app_MJ_printing:effects}
Each AM technique presents diverse sources of defects that have an impact on the final printed part, either in the geometry or in the mechanical properties. In general, the most common manufacturing parameters to be considered during the majority of the printing process are build orientation, layer thickness, infill patterns, raster angle, density, temperature, printing speed, nozzle diameter, and supports. In the case of vat photopolymerization and MJ, the UV curing parameter is additionally taken into account. The post-processing stage varies among the different technologies, and it involves support removal, which can be done by hand, water jetting, or chemical methods.
Geometrical defects on the final printed part are attained in the AM process. Imperfections, high levels of roughness, or residual porosity tend to reduce the mechanical performances of the fabricated parts due to increments of stress and strain concentrations~\citep{boniotti2019analysis}. Offset nodes, strut waviness or even incomplete structures (e.g. not printed struts) are typical imperfections found on micro-lattice structures induced by the AM process~\citep{zhang2020design}. The printing orientation leads to significant nodal volumes and the presence of slenderness on the struts of SLA micro-lattices~\citep{kudo2019compressive}.
The existence of anisotropic properties is another consequence of the additive process, it is caused by the layers' build-up operations and it varies depending on the technique. In particular, FDM or SLS produce parts with a high level of anisotropy ($\approx$~50\% and~10\%. respectively), while it is lower for SLA ($\approx$~1\%) and MJ ($\approx$~2\%)~\citep{kazmer2017three}. Despite existing studies that analyze the printing defects on massively used technologies like material extrusion, SLM, EBM, SLA, further research is still required to find accurate models. 
These outcomes and applied methodologies in photopolymerization techniques might be extended to other similar processes, like MJ. Thereby, more accurate mechanical performances of printed parts can be estimated.

\section{Cost estimation of architected lattices via MJ}\label{app_MJ_cost}
This case study aims to evaluate an approximated cost related to the MJ fabrication of an octet truss unit cell (30x30x30~mm with ligaments diameter of 4.0~mm and real volume of 5.33~cm$^{3}$) and the corresponding three-dimensional 4x4x4 lattice arrangement (with real volume of 341.30~cm$^{3}$). The currency used for the calculations was USD (\$) and prices without VAT. A Polyjet printer, Stratasys J750, was employed to estimate the printing time ($t_{print}$) based on the volume of the geometry to be fabricated, printing modes, and type of surface finishing by means of the complementary GrabCAD Print software. Similarly, the approximated model and support material weights were obtained, as seen in Table~\ref{Table:2}. The referential material prices were taken from Italian market quotations for late 2021. For example, in the case of VeroYellow RGD 836, \mbox{$M_{cost}=317.80$}~\$/kg and for SUP706B, \mbox{$M_{cost}=114.30$}~\$/kg. In addition, a waste material factor $\omega$=1.76~\citep{baumers2015modeling} was also introduced to include the possible resin that tends to be accumulated on the level roller after each print heads passage. 
In general, it is difficult to predict the amount of energy employed during the printing operations, and it does not represent a predominant value in the final market cost~\citep{shapira2017next}. 
For the scope of our estimation by using a printer Stratasys J750, the total energy consumption $E_{total}$ invested during the printing stage (preliminary and build-up process time), was calculated with the expression  \mbox{$E_{total}=E_{start}+(E_{process}\cdot t_{print})$.} 
Where, the considered values related to power-on, warm-up, and pre-printing stages \mbox{($E_{start}=0.10$~MJ)} and the printing process itself (\mbox{$E_{process}=533.1$~J/s}), were based on referential quantities from analogous predecessor printer models of the same Polyjet family (e.g. Objet Connex260~\citep{baumers2015modeling}, Eden 260v~\citep{kazmer2017three}). The energy price was referred to March 2022 suggested values ($E_{price}=0.0687$\$~/MJ). 
Suggested indirect cost $I_{cost}$, includes machine depreciation related to a Stratasys J750 printer, and it is equal to the expression \mbox{$I_{cost}= (MC\div t_{dep}\cdot u_{annual})\cdot t_{print}$}~\citep{li2017cost}. Where, the estimated  Machine Cost ($MC$) was $MC=264$~k\$, referred to Italian market quotations for late 2021. The considered depreciation time and annual utilization rate values were $t_{dep}=$8 years and \mbox{$u_{annual}=57\%$}, respectively~\citep{li2017cost}.
Labor cost ($L_{cost}$) can be calculated by \mbox{$L_{cost}= (S_{hour}\cdot t_{op})$}. For example, based on the salary per hour of a technician from the University of Trento ($S_{hour}=15$~\$/hour for the year 2022) and an approximated time fraction for printing operations of $t_{op}=1$~hour. 

\onecolumn
\begin{multicols}{2}
\noindent Thus, a value of $L_{cost}=15~\$$ can be added to the Total Cost ($TC$) from Table~\ref{Table:2}, which is provided by raw prices of materials and machine for all the cases, either single octet unit cell, lattice and both printing processes, Material Jetting or FDM. Moreover, an FDM cost estimation was also provided, by considering the same case study samples and assessment criteria. The following parameters were used: 3D printer Prusa i3 MK3S with the printing settings in normal mode of Ultra-detail (UD) with 0.05~mm for layers' thickness and Quality with 0.15~mm for layers' thickness, average energy process consumption ($E_{process}=100$~J/s) and machine cost $MC=999$~\$~\citep{Prusa}. The approximated printing time for the specimens fabrication, as well as the materials weight consumption, were calculated by the software Prusa Slicer. The selected filament was Prusament PC Blend Jet Black (diameter=1.75~mm) with \mbox{$M_{cost}=51.11$}~\$/kg~\citep{PrusaPCblend}, and the soluble supports Verbatim BVOH with \mbox{$M_{cost}=158.68$}~\$/kg~\citep{PrusaSupport}.
\end{multicols}

\section{Constitutive models for material jetting architected lattices}\label{summary:CM:AL}
\begingroup
\renewcommand\arraystretch{1.5}
\begin{longtable}{*1{>{\centering\arraybackslash}m{0.2\textwidth}}*2{>{\centering\arraybackslash}m{0.15\textwidth}}*3{>{\centering\arraybackslash}m{0.25\textwidth}}}
  \caption{Constitutive models adopted in the FE analysis to simulate the mechanical response of MJ architected lattices depending on the printing material.}\label{tab:LatCModels}\\
  \midrule[\heavyrulewidth] \textbf{Research work} & \textbf{Base Materials} & \textbf{Description} & \textbf{Constitutive model}\\\hline
  \endfirsthead
  \multicolumn{4}{r}{{\textit{Continued from previous page}}} \\ 
  \midrule[\heavyrulewidth] \textbf{Research work} & \textbf{Base Materials} & \textbf{Description}  & \textbf{Numerical model} \\\hline
  \endhead
  \multicolumn{4}{r}{{\textit{Continued on next page}}} \\ 
  \endfoot
  \multicolumn{4}{r}{{\textit{Table end}}} \\
  \endlastfoot
  
 Mohammadi, M.R et al. (2020)~\citep{mohammadi2020hybrid} & VisiJet M3 Crystal & Rigid & Bilinear perfectly  plastic \\ 
 Kumar, S. et al. (2019)~\citep{kumar2019tunable} & VeroWhite & Rigid  & Combined Drucker-Praguer and Rankine surface \\
 Li, T. et al. (2021)~\citep{li2021additive} & VisiJet M3 Black & Rigid  & Linear-perfectly plastic \\
 Zhang, Z. et al. (2021)~\citep{zhang2021harnessing}   &  VeroWhitePlus & Rigid  & Maxwell-viscoelastic \\
 Cui, S. et al. (2018)~\citep{cui2018mechanical} & VeroWhitePlus & Rigid & Desayi and Krishnan Plastic-Damage \\
 Wagner, M. et al. (2017)~\citep{wagner2017large} & VeroWhitePlus RGD835 & Rigid & Viscoelastic- Prony series + Williams-Landel-Ferry shift factors \\
 Li, T. et al. (2018)~\citep{li2018exploiting}  & VeroWhite & Rigid  & Elastic-viscoplastic \\
 Albertini, F. et al. (2021)~\citep{albertini2021experimental}  & VeroWhite & Rigid  & Piobert-Luders elasto-plastic with isotropic non-linear hardening \\
 Zhang, P. et al. (2016)~\citep{zhang2016transversely}  & VeroWhitePlus & Rigid  & Transversely-Isotropic Hyperelastic-viscoplastic with Tsai-Wu failure criterion \\
 Chen, Y. et al. (2018)~\citep{chen20183d}  & VeroWhite  & Rigid  & Elasto-plastic \\ 
 Bossart, A. et al. (2021)~\citep{bossart2021oligomodal}  & VeroWhitePlus & Rigid & Linear elastic \\\hline
 Fernandes, M.C. et al. (2021)~\citep{fernandes2021mechanically}  & FLX9795-DM & Flexible & Neo-Hookean \\
 Jamshidian, M. et al. (2020)~\citep{jamshidian2020multiscale}   & Tango+ & Flexible & Hyperfoam\\
 Weeger, O. et al. (2019)~\citep{weeger2019digital} & TangoBlack+, Tango+  & Flexible & Weeger et al. model - Non linear hyperelastic \\
 Jiang, H. et al. (2020)~\citep{jiang20203d} & FLX95595-DM & Flexible & Arruda-Boyce \\
 Chen, Y. et al. (2017)~\citep{chen2017lattice}  & FLX9795-DM & Flexible & Arruda-Boyce \\
 Janbaz, S. et al. (2018)~\citep{janbaz2018multimaterial}  & DM: Vero and Tango+ A =60, Agilus, Tango+ & Flexible & Arruda-Boyce \\
 Li, T. et al. (2018)~\citep{li2018exploiting}  & Tango+ & Flexible  & Arruda Boyce \\
 Bossart, A. et al. (2021)~\citep{bossart2021oligomodal}  & Agilus30 & Flexible   & Arruda-Boyce *Viscoelasticity: Three-term Prony series \\
 Albertini, F. et al. (2021)~\citep{albertini2021experimental}  & TangoBlack & Flexible  & Arruda Boyce \\
 Mirzali, M.J. et al. (2020)~\citep{mirzaali2020non}  & Agilus30 Black (FLX985) & Flexible  & Neo-Hookean \\
 
\end{longtable}
\endgroup

 \newpage
\section{Relevant aspects of the reviewed material jetting architected lattices}\label{summary:AL}
\begingroup
\renewcommand\arraystretch{1.3}
\begin{longtable}{*1{>{\centering\arraybackslash}m{0.12\textwidth}}*2{>{\centering\arraybackslash}m{0.11\textwidth}}*3{>{\centering\arraybackslash}m{0.17\textwidth}}}
  \caption{Summary of the relevant aspects characterizing the material jetting architected lattices discussed in this review such as (i) used 3D printer, (ii) chosen printing materials, (iii) considered unit cell type, and (iv) potential applications of the research results indicated by the respective authors. }\label{tab:SM}\\
  \midrule[\heavyrulewidth] \textbf{Research work} & \textbf{3D printer} & \textbf{Materials} & \textbf{Unit cell type}&\textbf{Outcomes} & \textbf{Applications} \\\hline
  \endfirsthead
  \multicolumn{6}{r}{{\textit{Continued from previous page}}} \\ 
  \midrule[\heavyrulewidth] \textbf{Research work} & \textbf{3D printer} & \textbf{Materials} & \textbf{Unit cell type}&\textbf{Outcomes} & \textbf{Applications} \\\hline
  \endhead
  \multicolumn{6}{r}{{\textit{Continued on next page}}} \\ 
  \endfoot
  \multicolumn{6}{r}{{\textit{Table end}}} \\
  \endlastfoot
  Lancea, C. et al. (2020)~\citep{lancea2020compressive}  & Connex 500 & VeroClear (RGD810) & Sphere-based with/without cylindrical stiffening elements & High compressive strength to mass ratio & Aerospace, Automotive, Machinery\\
  Sathishkumar, N. et al. (2019)~\citep{sathishkumar2020mechanical}  & Objet Eden 260V & VeroWhitePlus, SUP 705 & TPMS Schoen’s: FRD, Octo & TPMS compressive strength assessment & 3D printing, infills, Dampeners, Medical implants/scaffolds\\
 Mohammadi, M.R et al. (2020)~\citep{mohammadi2020hybrid} & Projet 3500 HD Max & VisiJet M3 Crystal, VisiJet S300 & Pentamodes &  Different mechanical properties due to pentamodes' midpoint variations & Metafluids: High stiffness in one direction and high compliance in others\\ 
 Abate, K.M. et al. (2020)~\citep{abate2020design} & Projet 3510 HDMax  & VisiJet M3 Crystal & Vintiles & Optimization of mechanical properties for high porosity lattices.  & Medical implants/scaffolds\\ 
 Kumar, S. et al. (2019)~\citep{kumar2019tunable} & Objet260  & VeroWhite, SUP705 & Honeycomb: Irregular hexagonal, Re-entrant, Chiral  & High energy absorption efficiency by tuning the gradient of wall thickness & Energy storage, Reconfigurable devices, Insulators, Dampeners\\ 
 Li, T. et al. (2021)~\citep{li2021additive} & Projet MJP 3600 & VisiJet M3 Black, VisiJet S300 & Semi-plate lattices:Simple cubic (SC), Face-centered cubic (FCC)  & Higher strength and energy storage levels than truss topologies, beyond Hashin-Shtrikman bounds. & Alternative to Metal foams\\ 
 Afshar, M. et al. (2018)~\citep{afshar2018compressive} & Objet EDEN & VeroGray Fullcure 850, SUP 705  & TPMS with graded porosity: P, D, IW-P & Porosity effects of TMPS scaffolds on plastic deformations and energy absorption. & Scaffolds for cells growth\\ 
 Kadkhodapour, J. et al. (2014)~\citep{kadkhodapour2014investigating}   & Objet Eden260  & VeroBlue FullCure 840 & TPMS: P and D & Assessment of the large deformation response of TPMS for scaffolds & Scaffolds for cells growth\\ 
 Alaboodi, A.S et al. (2018)~\citep{alaboodi2018experimental}  & ProJet 3500 3D & VisiJet M3 Crystal, VisiJet M3 White   & Cubic reticulated & Evaluation of mechanical and fracture behaviour of coloured and translucent porous lattices & Scaffolds for bones cells growth\\
 Alberdi, R. et al. (2020)~\citep{li2021additive}  & Objet 30 & VeroWhite & Combined: Face-centered cubic (FCC) and Body-centered cubic (BCC) & Tunable lattices with improved energy absorption & Impact and crash devices (with energy absorption through plastic dissipation)\\ 
 White, B.C. et al. (2021)~\citep{white2021interpenetrating}   & Objet J286 & VeroWhite & IPL combined: Face-centered cubic (FCC) and Dual-Rhombic Dodecahedron (RD) & Energy absorption six times higher than single separated lattices & Energy dissipation, Damage sensing, Vibration isolation \\
 Zhang, Z. et al. (2021)~\citep{zhang2021harnessing}   & Objet260 Connex3 & VeroWhitePlus & Square fractal patterns with three orders & Low bending stiffness but high structural integrity and improved shape recovery & Energy dissipation \\ 
 Johnson, L.K. et al. (2019)~\citep{johnson20183d}   & Objet500 Connex3 & Tango+ (FLX930), SUP706 & Cubic lattice with cylindrical struts & Stiffness values close to soft biological materials by varying geometrical features & Soft tissues\\ 
 Fernandes, M.C. et al. (2021)~\citep{fernandes2021mechanically}  & Objet500 Connex3 & FLX9795-DM & Square grid with diagonal bracings (deep-sea sponge inspired) & Nonlinear response and large deformations lattices & Energy absorption, Mitigation of acoustic and thermal waves\\ 
 Janbaz, S. et al. (2019)~\citep{janbaz2019ultra}  & Object 350 Connex3 & DM: Agilus + VeroMagenta, SUP706 & Four-Fold, Circular & Flexible Lattices driven by a buckling-mode controlled mechanism. High instability modes activated & Biomedical prosthetics, Soft robotics actuators (exoskeletons)\\ 
 Jamshidian, M. et al. (2020)~\citep{jamshidian2020multiscale}   & Stratasys J750  & Tango+ & Body-centered cubic (BCC) & Initial Model towards further nonlinear constitutive models (for lattices under large deformation)  & Soft lattices modelling \\ 
 Weeger, O. et al. (2019)~\citep{weeger2019digital} & Stratasys J750  & TangoBlack+, Tango+, SUP 706. Shoe sole: DM 2160 lattices+VeroPlus surface & Vintiles, Cross, Octet & New non-linear approach for soft lattices under large strains with practical applications & Soft robotics, Energy absorption, Fashion and personal protection\\ 
 Cui, S. et al. (2018)~\citep{cui2018mechanical} & - & VeroWhitePlus & Cubic with Sine-shaped struts & Variable NPR. Improved stiffness and fracture toughness than open cell foams. High Indentation resistance & Aerospace, Polymer seat cushions\\ 
 Jiang, H. et al. (2020)~\citep{jiang20203d} & Objet260 & FLX95595-DM & Sinusoidal beams  & Evaluation of mechanical properties of hyperelastic 3D tubular lattices & Biomedical Tissues, Stretchable electronics\\ 
 Mousanezhad, D. et al. (2015)~\citep{mousanezhad2015hierarchical} & Objet Eden260V  & TangoGray  & 2D hierarchical honeycomb & Evaluation of mechanical properties of Hierarchical honeycombs & Energy absorption, Tunable membrane filters, Acoustic dampeners\\ 
 Li, T. et al. (2017)~\citep{li2017harnessing} & Objet Connex260 & FLX9795-DM & Sinusoidal beams & Tunable NPR over large strains in all directions & Energy absorption, Tunable acoustics, Vibration control, Stretchable electronics, Soft robotics\\ 
 Wagner, M. et al. (2017)~\citep{wagner2017large} & Objet500 Connex3 & VeroWhitePlus RGD835 & Honeycombs: Re-entrant, Anti-tetrachiral & Active programmable auxetic honeycombs with area changes close to 200\% & Biomedical devices, Aerospace, Deployable structures, Self-assembly devices\\ 
 Wang, D. et al. (2020)~\citep{wang2020design}  & Objet J750 & VeroWhite & Horseshoe 3D lattice & Tunable Poisson’s ratio, programmable shape changes/deformations  & Soft robotics, Energy absorption, Biomedical devices\\ 
 Dong, L. et al. (2021)~\citep{dong2021design}  & Objet J750 & VeroWhite & Honeycombs with curled struts & Controllable shapes by geometry and temperature.  Variable Poisson’s ratio   & Morphing structures, Biomedical devices, Soft robotics, Shock absorbing materials\\ 
 Guseinov, R. et al. (2020)~\citep{guseinov2020programming}  & Stratasys J750 & Vero Pure White & Cell tessellation with stretched edge membranes & Morphing programmable shells considering shape evolution in time & Self-actuating shells, Shape-programmed robotic materials\\ 
 Xie, Y. et al. (2018)~\citep{xie2018acoustic}  & Stratasys J750 & - & 3D-cross shaped & 2.5D Luneburg lenses for 40 kHz airborne ultrasound.  & Control of airborne acoustic waves. Ultrasound sensors\\ 
 Peri, V. et al. (2019)~\citep{peri2019axial}  & Connex Objet500 & VeroWhitePlus. & Chiral & Phononic metamaterial with an induced axial field and non-local Weyl orbits presence.  Chiral Landau levels observed & Acoustic and electromagnetic metamaterials\\ 
 Vdovin, R. et al. (2017)~\citep{vdovin2017implementation}  & Objet Eden 350 & VeroWhite 830 & Thin-walled cells coupled by tubes &  High sound absorption efficiency (between 300 to 600Hz range) & Sound absorbers  \\ 
 Yi, J. et al. (2016)~\citep{yi20163d}  & Objet Eden260VS & - & Non-linear array of Steer beams &  Wave propagation control on a wide frequency range (from 8 to 12GHz) & Antennas, Microwave lenses, Airborne and trainborne devices for communication systems \\ 
 Chen, Y. et al. (2017)~\citep{chen2017broadband}  & Objet260 Connex3 & VeroWhite & Sinusoidal ligaments on Hexagonal, Square, Kagome, Triangular topologies   & Multiple and omnidirectional bandgaps tailored by the struts curvature & Wave filtering, Vibration isolation, Stretchable electronics, Programmable acoustic materials\\ 
 Chen, Y. et al. (2017)~\citep{chen2017lattice}  & Objet Connex260 & FLX9795-DM & Sinusoidal struts on several lattices. For NPR: Hexagonal, Square, Kagome, Triangular. For phononic band gaps: Square & Tunable Poisson’s ratio and wave propagation control & Energy absorption, Soft robotics, Vibration control, Tunable acoustic materials\\ 
 Lim, C.W. et al. (2021)~\citep{lim2021phononic}   & Objet30 & VeroWhite & External frame connected with central spherical/cubic masses & Scale independent meta-structures with elastic waves control in ultrawide frequency regions (improved band gaps width up to 160\%) & Vibration absorption and attenuation at wide frequencies in all directions. Underwater acoustic devices\\ \hline
 \textbf{ Multi-material} &  &  &  & \\ \hline
 Mirzali, M.J. et al. (2018)~\citep{mirzaali2018multi}  & Objet500 Connex3 & VeroCyan (RGD841) and Agilus30 Black (FLX985) & Honeycombs: Bow tie (60\textdegree), Rectangular (90\textdegree), Hexagonal (120\textdegree) & Elastic stiffness and Poisson Ratio (tunable in different directions), considering hard and soft phases &  Long term biomedical applications (implants). Soft Robotics \\ 
 Janbaz, S. et al. (2018)~\citep{janbaz2018multimaterial}  & Objet 350 Connex3 & DM Vero and Tango+ (A =60), Agilus, Tango+ & Squared-shaped elements linked with ligaments. * squared shapes inner core: circular, fourfold  & Instabilities and post buckling behavior controlled by voids distribution  & Stretchable electronics, Biomedical implants, Soft robotics\\ 
 Jiang, Y. et al. (2018)~\citep{jiang20183d}  & Objet Connex 260  & DM9760 and VeroWhite & Open cells with re-entrant angle chiral cores & Tunable stiffness, and Poisson’s ratio by geometrical features. Auxetic lattice made of rigid ribs and rubbery cell core  & Smart composites, Energy absorption foams, Sensors, Actuators, Biomedical devices\\
 Li, Y. et al. (2019)~\citep{li2019phononic}  & - & Tango+ and VeroWhite & Hoberman sphere with/ without soft or hard rods  & Multiple and tunable wide band gaps at low frequency ranges under Floquet Bloch boundary conditions & Sound insulation, Wave-guiding, Noise and vibration control\\ 
 Bossart, A. et al. (2021)~\citep{bossart2021oligomodal}  & Objet 500 Connex3  & VeroWhitePlus and Agilus30 & Tilings with soft squared hinges  & Controlled and constant zero-energy deformation modes. Tunable Poisson's ratio, negative under low compression and positive at high loading & Soft robotics. Energy absorption\\ 
 Al Ketan, O. et al. (2017)~\citep{al2017mechanical}   & Stratasys J750  & VeroWhite and Tango+ & TPMS: Diamond, I-WP (P and S), Gyroid, Fischer-Koch C(Y) & Mechanical properties assessment of TPMS based in interpenetrating composites (IPC) and their volume fractions & Vibration damping. Damage reduction\\ 
 Dalaq, A.S. et al. (2017)~\citep{dalaq2016mechanical}  & Objet260 Connex & Tango+ (FLX930), and Vero-Plus (RGD875) & TPMS: Schwarz (P, D, CLP), Schoen (I-WP), Neovious, Gyroid, Fischer-Koch S, Truss-IPC  & Mechanical properties improvements of IPC with a rubbery matrix reinforced by rigid TMPS lattices  & Damage tolerant devices. Vibration control\\ 
 Li, T. et al. (2018)~\citep{li2018exploiting}  & Objet Connex260 & VeroWhite and Tango+ & Reinforcing lattices: Auxetic(Re-entrant honeycomb, Chiral truss). No auxetic: (Honeycomb, Truss)   & Reinforced rubbery matrix with rigid auxetic lattices. Enhanced strength and energy absorption due to NPR effects & Energy absorption, Impact resistant devices  \\ 
 Albertini, F. et al. (2021)~\citep{albertini2021experimental}  & Objet Connex350 & VeroWhite and TangoBlack + & Hexaround and Warmuth & Enhanced mechanical properties and energy absorption of composite stiff lattices with hyperelastic material infill & Energy absorption\\ 
 Mirzali, M.J. et al. (2020)~\citep{mirzaali2020non}  & Objet500 Connex3 & VeroCyan (RGD841) and Agilus30 Black (FLX985) & Honeycombs: Bow tie (60\textdegree), Rectangular (90\textdegree), Hexagonal (120\textdegree) & Evaluation of non affinity auxetic lattices with two phases & Remarks for structural integrity in multi-materials design\\ 
 Al Ketan, O. et al. (2018)~\citep{al2018nature}  & Objet Connex 260 & VeroWhite and TangoGray & TMPS: Gyroid & Improved energy absorption of cellular composites made of graded TPMS with variable volume fraction & Energy absorption\\ 
 Mueller, J. et al. (2018)~\citep{mueller2018stepwise}  & Objet500 Connex3 & RGD525 and FLX9695-DM & Voronoi core & Improvements on Energy absorption of graded bending-dominated lattices up to 38 times with respect to single material counterparts & Impact protection devices, from helmets to vehicles\\ 
 Wang, K. et al. (2015)~\citep{wang2015designable}  & Connex350 & VeroWhitePlus and TangoBlack+ & 3D Re-entrant honeycomb & Evaluation of mechanical properties of auxetic lattices with flexible joints and rigid beams & Shock-resistant devices, Biomedical components, Smart materials\\ 
 Saxena, K. et al. (2017)~\citep{saxena20173d}  & Connex 260 & VeroWhite and Tango+ & 2D Re-entrant honeycomb & Assessment of Young's modulus and Poisson's ratio of multi-material auxetic lattices. Indentation resistance & Impact-protection devices.   \\ 
 Hedayati, R. et al. (2021)~\citep{hedayati20213d}  & Object500 Connex3 & Agilus 30 (60\%) and VeroBlack+ (40\%) & Hexagonal, Re-entrant (incorporated in several patterns along a cylindrical shape) & Analytical formulations to determine Poisson’s ratio and stiffness assessment & Smart materials, Soft-robotics, Exo-suits, Heavy-duty clothing, Energy damping, Soft actuators\\ 
 Lumpe, T.S. et al. (2021)~\citep{lumpe2021computational}  & Objet500 Connex3  & VeroWhitePlus and “High Temperature” (HT) & Squared and sine structure (2D and 3D) & Optimization of multimaterial morphing shape lattices under heat stimuli & Morphing components (e.g. active airfoils, or car panels)\\ 
 Wu, L. et al. (2016)~\citep{wu2016isotropic}  & Objet350 Connex2 & VeroWhitePlus (RGD835) and  Tango+ (FLX930) & 3D Anti-chiral & Tunable Negative thermal expansion coefficients, extendable to any lattice scale & Resistant components to thermal fatigue and cracking, (in fields like aerospace, architecture)\\ 
 Egan, P.F. et al. (2019)~\citep{egan2019mechanics}  & Objet500 Connex3  & MED610 bio-compatible polymer, SUP706 & BCC cage & Integrated design approach with manufacturing considerations (building orientation, post-processing and environmental exposure). Evaluation of mechanical properties & Scaffolds. Biomedical applications  \\ 
 Liu, W. et al. (2020)~\citep{liu2020maximizing}  & Objet30pro  & VeroWhitePlus (RGD835), SUP706  & BCC, BCC-Z, FCC Octets  & Snap fitted assembly method to reduce support material during lattice fabrication. Evaluation of compressive strength, elastic and inelastic buckling & Manufacturing procedures \\ 
 Mueller, J. et al. (2018)~\citep{mueller2018buckling}  & Objet500 Connex3 & VeroWhitePlus (RGD835), SUP705 & FCC, Kelvin cell & Assessment of Buckling, orientation and scaling significant effects on mechanical properties of lattices & Considerations for safety and high efficiency of 3D printed parts\\ 
 Zhang, P. et al. (2016)~\citep{zhang2016transversely}  & Objet260 Connex & VeroWhitePlus & Square-grid, Diamond & Printing effects like anisotropy  considered for mechanical properties assessment & Accurate deformation and failure prediction for polymeric 3D printed parts\\ 
 Chen, Y. et al. (2018)~\citep{chen20183d}  & Objet Connex260 & VeroWhite & Hierarchical hexagonal honeycombs & High energy dissipation and preservation of shape integrity under large deformation, by hierarchy levels and heat treatment & Energy absorption, Damage tolerant devices, Damping   \\ \hline
 
\end{longtable}
\endgroup
 
 \end{appendices}

\end{document}